# Using machine learning approaches to classify false-positives generated by static analysis tools - An empirical study using small datasets

LAKMAL DESHAPRIYA*

School of Computing, University of Otago, Dunedin, New Zealand, lakmal.deshapriya@postgrad.otago.ac.nz

SHERLOCK A. LICORISH

School of Computing, University of Otago, Dunedin, New Zealand, sherlock.licorish@otago.ac.nz

BRENDON J. WOODFORD

School of Computing, University of Otago, Dunedin, New Zealand, brendon.woodford@otago.ac.nz

Static Code Analysis (SCA) tools are vital for software development to reduce the cost and time required for manual code reviews. It was established in our previous study that the best tools reported in the software engineering community return high false-positive and false-negative rates when used for code analysis. Several approaches have been taken in existing studies to identify false-positive alarms generated by SCA tools, in helping to enhance the utility of these tools. However, existing studies reveal several limitations, chiefly among which is the tendency to focus on a limited number of alarm types or failing to evaluate a large breadth of Machine-Learning (ML) techniques comprehensively. This study has addressed this opportunity. We identified the best Code Representation Learning (CRL) techniques and ML algorithms for identifying false-positives, where results show that performance outcomes vary based on the characteristics of the training datasets and the goal of the model building. Given the evidence established in this study, we recommend further investigations targeting specific configurations of advanced CRLs and ML Algorithms, including simple ML models.

## 1 INTRODUCTION

Static Code Analysis (SCA) tools are essential to software development. They enable developers to detect software code quality issues, such as vulnerabilities and performance issues in source code. SCA tools analyse source code without execution, improving security and overall source code quality [1]. However, SCA tools are often plagued by high false-positive and false-negative rates, requiring developers to sift through numerous false alarms, diminishing their productivity and focus on resolving critical issues [1, 2].

Existing studies have explored various approaches to identify false-positive and false-negative warnings[1]. For instance, Pujar et al. [2] used their D2A dataset to train several Machine-Learning (ML) approaches using selected features of warnings and source code. As a result, four ML approaches were selected as best. Moreover, Yang et al. [3] used several ML approaches to train ML models[2]. They used the source code and the warning message as the input for ML model, and they found that ML approaches can provide good results. Moreover, Code Representation Learning techniques (CRLs) [3] and some ML approaches stand out from those existing approaches in terms of their performance towards classifying SCA warnings. In recent studies, Deep Learning (DL) approaches have been dominant where they were used to train ML models for classification tasks in various domains. However, according to Yedida et al. [4], simple ML approaches like the Random Forest (RF) algorithm perform better than DL approaches for SCA warning classification tasks.

* Corresponding author

[1] In this study, a warning denotes the identification of a code violation by a SCA tool.

[2] In this study, ML models denote the resulting trained models that combine embeddings with these algorithms, such as Word2Vec_BGRU and Code2Vec_RF.

While advancements in technologies such as CRLs and ML algorithms[3] have introduced sophisticated solutions, existing studies often focus on datasets with specific warning types using limited CRLs and ML algorithms [5-8]. Therefore, we have conducted experiments on two public datasets using two popular SCA tools which were identified as less prone to false-positives and false-negatives by a previous study [9]. From our experiments summarised in Section 4, we have identified the warning types that were mainly prone to false-positives. Extending our previous study, in the current work, we conducted an empirical investigation to explore the best CRLs and ML algorithms for identifying false-positives generated by PMD and SonarQube. Our study identifies performance patterns in ML models across various warnings using a relatively small dataset compared to existing studies. While it is challenging for most researchers to find a large, annotated dataset for their experiments, this study demonstrates that the software engineering industry can achieve good performance even with small datasets, provided the dataset is of high quality and the right ML models are employed.

In Section 3, we have identified several drawbacks in previous studies, including failure to evaluate ML algorithms and CRLs that have emerged as the most effective in existing related studies, advanced and novel ML approaches that were introduced recently were not evaluated to classify SCA warnings, and most studies have experimented on limited code violation types. Addressing these issues contribute significantly to the body of evidence aimed at evaluating SCA tools.

In fact, existing studies evaluated specific CRLs and ML algorithms to find false-positive warnings generated by SCA tools [5, 10-12], however, ML algorithms like BiLSTM have not been studied for SCA warning classification. Therefore, it is necessary to benchmark these technologies against those previously evaluated. In evaluating CRLs and ML algorithms, we found four CRLs and six ML algorithms produced the best results [3, 8, 12-14]. However, none of the existing studies evaluate all the technologies, in performing a detailed comparative evaluation. Therefore, it is worth considering the selected CRLs and ML algorithms that have proven best in existing studies for classifying SCA warnings in a comprehensive study to identify false-positives. This study provides an evaluation of the performance and behavioural patterns of wide range of CRLs when used for modelling across several ML algorithms for multiple datasets. We created four warning datasets, each with distinct features. Our analysis started with two different source code datasets: the Stack Overflow dataset, which includes code snippets, and the Apache Tomcat dataset, which contains complete source code files. We then used two SCA tools, including PMD and SonarQube, to analyse each source code dataset. This process produced four SCA warning datasets. Each warning dataset has unique features based on the type of source code (snippets versus full-length files) and the specific analysis capabilities of each SCA tool.

In performing our evaluation and addressing the need for comprehensive assessment [3, 8, 12-14], the first Research Question (RQ1) was developed to identify the impact of different CRLs on the performance of the ML algorithm in classifying false-positive SCA warnings. Source codes and SCA warning messages were embedded using CRLs, and ML algorithms were used to build ML models.

**RQ1: How does the performance of ML algorithms vary with the use of different CRLs in SCA warning classification?**

As outlined in Section 2, the performance of ML models in identifying false-positive warnings depends on the capabilities of the algorithms employed. While we have identified the most effective ML algorithms used in previous studies [3, 5, 8, 11-14], none of these studies evaluated a wide breadth of these ML algorithms in a single investigation. It is essential to conduct a comprehensive evaluation of the best-performing ML algorithms to explore how they perform under varying contexts. Thus, the second Research Question (RQ2) was formulated to address this gap.

---

[3] In this study, ML algorithms refer to the methods used to classify SCA warnings, such as BGRU, GRU, and RF.

**RQ2: How does SCA warning classification performance vary with the use of different ML algorithms?**

The performance of the ML model depends on the practical purpose of the model building [15], which is further discussed in the Section 2. We identified four possible purposes for building models in classifying SCA warnings and determined the most appropriate performance metrics to evaluate ML models for each purpose. There can also be variations in performance patterns depending on the metrics used for evaluation [15], which needs consideration. To address these gaps, the third research question has been formulated.

**RQ3: How does SCA warning classification performance vary depending on the purpose of building ML models?**

For our study, we utilised four warning datasets with different characteristics as explained in the Section 4. For RQ1 and RQ2, the evaluation was conducted on all datasets collectively to gain a general overview of the performance trends of CRLs and ML algorithms. However, variations in performance patterns may occur in ML models depending on the characteristics of the trained datasets. Therefore, to examine how the choice of dataset influences the performance of the selected machine learning models, the fourth research question has been formulated.

**RQ4: How does SCA warning classification performance vary with the use of different datasets?**

As noted above, this study used four datasets to answer four research questions. The datasets are summarised in Section 4, and the methodology for model building is documented in Section 5. In this regard, the main contributions of this study are as follows.

- Our evidence provides an opportunity for the community to select the best CRLs and ML algorithms from existing studies, and our datasets provide benchmarks data for future studies to use.
- We demonstrate that ML models can train with small datasets when they are of high quality, providing suggestions for those faced with data scarcity issues.
- We introduce multiple program-slicing approaches by adapting existing approaches for false-positive detection experiments. Here we provide support for software fault localisation efforts more generally.
- We provide a replication package to support follow-up investigations. Our package includes detailed results from modelling along with optimised hyperparameters.

The rest of the study is organised as follows. Section 2 discusses the background of the study. Section 3 reviews existing studies related to building ML models to classify SCA warnings and Section 4 outlines the datasets we used for our study. Section 5 documents the methodology followed in our study and Section 6 summarises the results generated in our study. We then discuss our results and their implications in Section 7. In Section 8, we outline potential threats to the validity of our outcomes. Section 9 concludes the work and outlines potential areas for future research. Moreover, we provide a replication package that includes the appendices and training datasets used in this study [16].

# 2 BACKGROUND

## 2.1 Performance metrics used in this study

Several performance metrics were used to evaluate ML models experimented with in this study, including MCC, F1-Score, Balanced Accuracy and Computational Complexity. **The Matthews Correlation Coefficient (MCC)** is a balanced metric used to evaluate the performance of binary classifiers, particularly when dealing with imbalanced datasets. It is frequently utilised to assess binary classification models and is computed using the following formula [17]:

$$MCC = \frac{TP \times TN - FP \times FN}{\sqrt{(TP + FP)(TP + FN)(TN + FP)(TN + FN)}}$$

Where,
True Positives (TP) = Number of correctly predicted positive instances.
True Negatives (TN) = Number of correctly predicted negative instances.
False Positives (FP) = Number of incorrectly predicted positive instances.
False Negatives (FN) = Number of incorrectly predicted negative instances.

The MCC is a key metric for assessing the performance of binary classification models. It accounts for all four categories of the confusion matrix: true positives, true negatives, false-positives, and false-negatives. This comprehensive approach makes MCC especially valuable when working with imbalanced datasets, where metrics like accuracy can be deceptive. A MCC score of +1 indicates perfect predictions, zero signifies performance no better than random guessing, and -1 represents complete disagreement between predictions and actual outcomes. In summary, MCC is a reliable indicator of model effectiveness, making it a crucial tool for researchers and practitioners in machine learning and statistical analysis [18].

The **F1-Score** is a statistical measure used to evaluate the performance of a binary classification model. It is the harmonic mean of Precision and Recall, providing a balance between the two metrics. Here, **Precision** measures the accuracy of positive predictions, while **Recall** (or sensitivity) measures the ability of the model to identify all relevant instances [19, 20]. The formula for F1-Score is:

$$\text{F1} - \text{Score} = 2 \times \frac{Precision \times Recall}{Precision + Recall}$$

Where,

$$Precision = \frac{TP}{TP + FP}$$

$$Recall/Sensivity = \frac{TP}{TP + FN}$$

The F1-Score ranges from 0 to 1, where 1 indicates perfect Precision and Recall, and 0 indicates the poorest performance. It is beneficial as a measure when the class distribution is imbalanced, and the evaluation needs to consider both false-positives and false-negatives [20].

**Balanced Accuracy** is a metric used to assess the performance of a classification model, particularly when the classes are unequally distributed. It provides a more meaningful assessment than traditional accuracy, which can be misleading if one class dominates. Balanced Accuracy is calculated by averaging Sensitivity and Specificity. The formula is:

$$Balanced\ Accuracy = \frac{Sensitivity + Specificity}{2}$$

Where,

$$Specificity = \frac{TN}{TN + FP}$$

In this formula, Sensitivity (Recall) is the ratio of true positives to the sum of true-positives and false-negatives, indicating how effectively the model detects the positive class. Specificity is the ratio of true negatives to the sum of true-negatives and false-positives, assessing how well the model identifies the negative class. In summary, Balanced Accuracy considers both the true positive and true negative rates, making it particularly useful for evaluating models on imbalanced datasets where one class may dominate the other [18]. By treating both classes equally, Balanced Accuracy helps ensure that the model performs well across all categories, rather than just on the majority class [18].

**Computational Complexity** in ML model building refers to the amount of computational resources required to train and evaluate a model. This includes time complexity (how an algorithm's running time scales with the input data's size) and space complexity (the amount of memory needed to store data and intermediate computations). It involves analysing how different algorithms might perform as the dataset size or the model's complexity increases [21]. For example, some algorithms perform well with small datasets but become inefficient as datasets grow larger. Understanding Computational Complexity aids in choosing suitable algorithms and tools for model development, particularly when dealing with large datasets or real-time systems. It helps balance model performance with the limitations of available computational resources. In this study, we constructed each model in a similar environment (refer to Section 5.3) and recorded the time required for training. The information regarding Computational Complexity can be used when selecting ML models to classify SCA warnings and identify false positive warnings. In our study, we measured the time taken to train each ML model, excluding the time required to create the numerical vector by embedding the source code and SCA warning description. Consequently, models with lower computational cost would be preferred.

## 2.2 Purposes of building ML models in classifying SCA warnings

We conducted a comprehensive literature study and summarised the outcome in Section 3. It is observed that existing studies did not consider the potential purposes of building ML models to classify SCA warnings. Instead, existing studies have tried to optimise ML models with random performance metrics without focusing on these purposes [10-12, 22]. Therefore, we identified four main goals for this study in classifying warnings generated by SCA tools, and suitable performance metrics to measure these goals.

The first model-building goal would be to maximise the overall model performance. Therefore, the model will accurately identify actual and false-positive warnings as much as possible, reducing incorrect predictions. For this scenario, metrics such as accuracy provide a good understanding of the performance of the ML models and how well all the warnings are classified. However, the datasets used in this study are biased towards actual or false-positive warnings. Therefore, the accuracy performance metric provides less reliable insights regarding model performance. Consequently, we used the MCC performance metric to identify the effectiveness of ML models, as in this study, we performed binary classifications for the model building.

The second goal of the model building is to detect false-positive warnings as much as possible to minimise false warnings. This helps to reduce developers' frustration with identifying false-positive warnings. However, optimising ML models to identify as many false-positive warnings as possible should not result in actual warnings being misclassified as false-positive warnings. In our dataset, false-positive warnings are labelled as "1" and actual warnings are labelled as "0". Therefore, we need to optimise ML models using a performance metric that maximises true positives and minimises false-positives and false-negatives. In this scenario, reducing false-positives is mandatory, as it will classify actual warnings as non-actionable warnings, and developers will eliminate these warnings without further investigation. In contrast, reducing false-negatives is also essential, as it will classify false-positive warnings as actual warnings. If the count of this case is higher, further investigation of false-negatives will waste the developer's time.

However, minimising false-negatives is less critical compared to minimising false-positives. Considering all these requirements, the F1-Score performance metric is the most suitable for the second goal of model building. As noted above, F1-Score is the harmonic mean of Precision and Recall performance metrics [22]. When the F1-Score equals 1.0, all the false-positive warnings are accurately classified as false-positive warnings, with no misses and no errors. Therefore, in the ideal situation (F1-Score = 1.0), developers can disregard all the identified false-positive warnings and focus on the remaining warnings. Hence, if the goal of model building is to optimise the model with the second goal, the objective is to find the best model (s) with the best F1-Score. Therefore, developers can eliminate all the warnings classified as positives. However, the eliminated warnings can include some actionable warnings, especially when the F1-Score is not perfect.

The third goal of the model building is to identify the maximum possible actual warnings with a minimum number of false-positive warnings, where the loss of actual warnings is minimised. However, false positive warnings may still occur. Considering the confusion matrix, the goal is to maximise true negatives while minimising false-positives and false negatives. Similar to the second goal, minimising false-positives is more important than reducing false-negatives. In this context, Balanced Accuracy can be utilised as a performance metric, where both the True Positive Rate (TPR) and the True Negative Rate (TNR) are taken into account [18]. Usage of TPR and TNR makes Balanced Accuracy particularly useful in situations where the classes are imbalanced. In the ideal situation where the Balanced Accuracy value equals 1.0, all actual warnings are identified by the ML model (s). That said, developers may reduce the time spent identifying false-positive warnings. On the other hand, the same goal can be achieved by identifying false-positive warnings as determined in the second goal using F1-Score performance metrics and removing those from the whole set of warnings. However, while finding ML models with perfect F1-Score values is challenging, striving for ML models with the best Balanced Accuracy value would be a practically feasible solution.

The fourth goal would be to reduce Computational Complexity. This will reduce the model's training time, allowing us to fine-tune ML models quickly. Consequently, models will generate higher performance. For this goal, Computational Complexity performance metrics can be used [21]. All the goals discussed in this section are summarised in Table 1. Additionally, we summarise the results generated from this study in Section 6, in terms of the performance metrics discussed in Section 2.1. We discussed these results in Section 7.

Table 1 Goals for model building

| Goal ID | Purpose of model building | Metric used |
|---|---|---|
| G1 | Maximise the overall model performance | MCC |
| G2 | Correctly detect false-positive warnings as much as possible to minimise false warnings | F1-Score |
| G3 | Correctly identify the maximum possible actual warnings with minimum false-positive warnings | Balanced Accuracy |
| G4 | Faster training and evaluation of ML models with limited computational resources, so that the models can be fine-tuned for better performance | Computational Complexity |

## 2.3 AST graphs and data dependency graphs

Only a slice of source code is relevant for the warning generated by the SCA tool, and identifying the relevant slice is critical for the performance of the classification model [11, 12, 23, 24]. To extract the relevant slice of the source code, several approaches are used [11, 12, 23, 24]. An Abstract Syntax Tree (AST) is a tree-like representation of software source code that illustrates its structure [25]. Each node in this tree signifies a particular construct (classes, methods, variables, etc.) found in the source code. During the conversion of source code into an AST, only the structural and content-related details of the source code are transferred, while any additional details, such as comments, are discarded. Information that gets transferred to the AST includes:

- The variable types and their declaration locations
- The order and definition of executable statements
- The left and right components of binary operations
- Identifiers and their assigned values

In this study, AST graphs are employed to extract code slices from source code where the relevant lines are dispersed throughout the source code, and the relationship between those lines depends on their structure.

In contrast, a data dependency graph is a directed graph illustrating the dependencies between data items in source code [26]. For instance, within source code, for a variable definition with an initial value, the modification of that variable and the utilisation of the variable's value present data dependencies. In this study, data dependency graphs are employed to extract code slices from source code, where the relevant lines are scattered throughout the code, and the relationship between these lines relies on their data or identifiers/variables.

# 3 RELATED WORK

## 3.1 CRLs and ML algorithms used to classify SCA warnings

SCA tools generate a high false-positive rate, which wastes time developers' use for finding and eliminating false alarms [1]. Therefore, efforts have been invested towards automating the task of eliminating false warnings and identifying actual warnings, emphasising the importance of classifying SCA alarms into true and false-positives [6, 13]. Existing studies have taken different approaches, such as using extracted features as input to train ML models, hybrid approaches requiring human involvement, ranking the warning based on the likelihood of the alarm being true positive and using ML approaches on CRLs [11, 27-29]. We review the studies in this regard in this section.

### *3.1.1 ML model building using extracted features*

Previous studies have utilised features extracted from SCA alarms and/or source code rather than directly incorporating source code as a training feature for ML models. Despite certain limitations, including the challenges in extracting all relevant features and uncertainty about whether the extracted features adequately represent the source code, using these features as input simplifies ML models, making it easier to fine-tune them [4, 30]. Alikhashashneh et al. [31] evaluated four ML approaches, SVM, KNN, RF, and Repeated Incremental Pruning to Produce Error Reduction (RIPPER) to classify warnings generated by SCA tools. They used software engineering metrics (source code metrics) like "codeline_count" and "Cyclomatic" (Cyclomatic complexity) to train the models, where RF was seen to be the best performer on average compared to the other examined ML techniques. According to their study, the most critical software engineering metrics were the "CountInput", "Knots", "CountOutput", "CountPath", and "Cyclomatic"; and these metrics were essential to predict the behaviour of SCA tools against a given piece of source code. Moreover, they found that reducing source code complexity, coupling, and usage of global variables in source code reduce false-positive and false-negative warnings generated by SCA tools. Cheirdari and Karabatis [32] also experimented with classifying warnings and found that features like "Author", "vulnerability ID", and "CWE ID" are significant when classifying SCA warnings. Overall, outcomes here show that Alikhashashneh et al. [31] found technical source code features to be correlated with SCA warning classification, while Cheirdari and Karabatis [32] identified personal identifiers to be more significant.

Berman et al. [33] experimented with using active learning to identify false-positives in SCA warnings while reducing the need for a large, labelled dataset. They used RF and lasso logistic regression with three active learning algorithms (uncertainty, query by committee, and random selection). Random Forest produced the best accuracy, while Query by Committee active learning returned the best results in terms of requiring the minimum number of labelled training examples to classify SCA alerts correctly. However, these authors did not use source code directly to train the model. Instead, they used features extracted from the alerts generated by static code analysis tools, which may not fully represent source code [4, 30], invoking curiosity about the potential for other approaches to represent SCA warnings (e.g., CRLs).

Qing et al. [34] evaluated six ML models (Naïve Bayes (NBayes), Logistic Regression, K-Nearest Neighbour (KNN), RF, Decision Tree (DT) and Support Vector Machine (SVM)) to classify SCA warning in a dataset created by Wang et al.[35] using 91 warning features, where it was found that out of those features, only 25 were relevant. This study revealed that Logistic Regression performed best in terms of false-positive rate when classifying SCA warnings. Yang et al. [27] used the "Golden Set" features suggested by Wang et al. [35] to train several ML models to classify SCA warnings. Wang et al. [35] identified 23 features, such as warning context in method and warning context in file, related to software source code and SCA warnings, which can be used to classify the SCA warnings into actionable/non-actionable warnings. They referred to these features as the "Golden Set". The method presented by Levina and Bickel [36] was used to reduce the feature dimensionality of the SCA warnings. They used several ML algorithms to train ML models for the classification task, including Deep Neural Network (DNN), Convolutional Neural Network (CNN), DNN_weighted (similar to the DNN with a weighted strategy), RF, DT, and Linear SVM. In this latter study, Linear SVM performed best for the dataset used. Additionally, the authors recommended Linear SVM for classifying SCA warnings if the dataset is low-dimensional.

Flynn et al. [37] proposed a novel approach to use a test suit (Juliet C/C++ test suit) as a dataset for classifying warnings generated by SCA tools. They extracted metadata from the dataset and used RF (H2O.ai implementation of RF), Lasso (the glmnet implementation of lasso-regularized logistic regression), LightGBM and XGBoost for the classification. In this work, LightGBM generated the best results in terms of performance metrics such as Accuracy, Recall , and AUPRC (Area Under the Precision–Recall Curve) for classifying SCA warnings.

Yang et al. [38] proposed a human-in-the-loop active learning [39] approach to identify actionable warnings from those generated by SCA tools. They used features extracted from both software source code and SCA warnings. Additionally, they used active learning with SVM, RF, and Decision Tree algorithms for warning classification and Active learning with SVM produced the best results in terms of AUC (Area Under the Curve).

Kharkar et al. [40] proposed a Transformer-based ML approach to identify false-positives generated from analytic bug detectors. GPT-C produced a 17.5% Precision improvement on top of the baseline for null pointer warnings and demonstrated generalizability to other bug types like resource leaks. Experiments used feature-based and neural models, including DeepInferEnhance (a fine-tuned CodeBERTa model) and GPT-C (a pre-trained generative model based on GPT-2). Results showed that GPT-C achieves the highest Precision improvements, while DeepInferEnhance offers better Recall, showcasing a trade-off between the models.

Kang et al. [30] used SVM and K-Nearest Neighbours (KNN to classify SCA warnings using "Golden Set" features proposed by Wang et al. [35]. They found that SVM is better for the task. However, they found that existing Golden features underperformed with ML models, highlighting the significant limitations in existing approaches. Therefore, according to Kang et al. [30], the Golden features are not suitable for SCA warning classification tasks. Yedida et al. [4] extend the work done by Kang et al. [30] where they identified that global treatments perform worse than localised treatments for complex input data. Therefore, the authors introduced four local treatments and showed that they improved the results generated by Kang et al. [30]. While performing localised treatments may improve the ability to identify patterns in source code, recent CRL technologies also find patterns efficiently [41-43]. In fact, it was noted that training using older technologies, like feedforward networks, produce superior performance to more recent DL and CodeBERT models [41-43]. Simpler models may also run faster, allowing researchers the opportunity at fine-tuning more rapidly.

Pujar et al. [2] experimented with several ML approaches to classify their SCA warning dataset – D2A. They labelled the dataset using a novel approach called Differential Dataset Analysis and made the dataset publicly available. Using hand-crafted features as the input, they experimented with 13 well-known ML models, namely XGBoost [44], LightGBM [45], CatBoost [46] and the scikit-learn [47] implementations of RF, DTs, Extra Trees[4], K-means clustering, Gradient Boosting, Ada Boost, Linear Classifiers, Gaussian NBayes, Multinomial NBayes, and Complement NBayes. Based on the performance of those ML algorithms in terms of their AUC and F1-Score, the four best models – RF, Extra Trees, LightGBM, and CatBoost were selected for further analysis with Voting and Stacking ensembles.

The studies above show that training on features extracted from SCA warnings and/or software source code produced promising results despite limitations identified by Kang et al. [30] and Yedida et al. [4]. According to Kang et al. [30] and Yedida et al. [4], the extracted features may not fully represent source code and SCA warnings to the ML models. To address these issues, researchers have employed CRLs to represent source code to train ML models for classifying SCA warnings [42, 43, 48]. Usage of CRLs to represent source code eliminates effort required to extract features from both source code and SCA warnings and enables the utilisation of the latest advanced CRLs, which may represent relationships within source code and SCA warning message effectively [41-43].

### *3.1.2 ML models that used CRLs to represent source code*

CRLs were used to represent source code and SCA warnings to ML models, and reduce the effort required to extract features from source code and warning messages [41-43]. Koc et al.

[4] A machine learning classification algorithm, which is an ensembled method that builds multiple decision trees and combines their predictions. https://scikit-learn.org/stable/modules/generated/sklearn.ensemble.ExtraTreesClassifier.html

[49] used NBayes and Long Short-Term Memory (LSTM) ML models to classify SCA warnings using a balanced dataset. They extracted relevant slices of source code to feed the models as input, where LSTM produced the best results in terms of Precision, Recall and Accuracy. Lee et al. [10] used a CNN ML model on 10k historic static analysis alarms generated by six static analysis alarm types, which developers annotated. The dataset had about 10% of false-positive warnings, where the average Precision across all six checkers was 79.72%, showing that CNN models may achieve good performance for this task. Code lines were extracted from the code as lexical tokens and tokenised to feed those tokens to the Word2Vec embedding technique (using a skip-gram model) that was trained using source code. A CNN model was then trained on the vector resulting in the outcomes. However, the authors also reported trivial overfitting, correlation between model quality and data quality and difficulty in integrating the classifier into the Continues Integration (CI) pipeline. While these outcomes are interesting, model loss graphs and ROC graphs could have provided more concrete tests of the degree to which the models were overfitted. Additionally, there was need to use other datasets to help with supporting the inference reported.

Koc et al. [11] experimented with four families of ML approaches (Hand-engineered features, Bag of words, Recurrent Neural Networks, and Graph Neural Networks) to classify static code analysis warnings generated by three SCA tools. They used Word2Vec as the CRL technique and a class-balanced dataset for model training. In this study, LSTM outperformed the other ML models. Gated Graph Neural Networks (GGNNs) also generated promising results. The authors used two methods to represent the code: program splicing and AST. The authors experimented with transformations on source code to train with ML models and found that LSTM with "Extracting English Words from Identifiers" transformation produced the best results. However, this may not be necessary in all such scenarios as contextual CRLs such as CodeBERT is able to find relationships between words and return word vectors based on the context [3].

Tanwar et al. [12] used an Ensemble Learning model with XGBoost, RF, and Neural Network to classify static code analysis alarms. They used historic user commits as the dataset, which had approximately 12.17% of false-positive alarms in the training and validation datasets. Moreover, the model was setup in the development pipeline, and eventually, the developers labelled the alarms. Additionally, a pre-trained Code2Vec model was used for code embedding. Nagaraj et al. [5] used four ML models to classify the warnings generated by an SCA tool into true positives and false-positives. They used a Word2Vec word embedding model trained on their corpus of source code. They evaluated RF, SVM, XGBoost, and an Ensemble ML model, finding that the RF model generated the best results in terms of model Accuracy. This study demonstrates that even simple models are capable of identifying false-positive warnings.

Hegedűs and Ferenc [14] used Word2Vec as a CRL, which was trained using a public dataset to train Code2Vec. They evaluated DTs, RF, NBayes and Neural Network (NN) ML approaches with 224,484 real-world warnings, which had about 79% false positive warnings from 160 SonarQube rules. They found that the RF algorithm produced the best results in terms of Accuracy, F1-Score, Precision, and MCC values. Nguyen et al. [6] presented a novel approach to identify and reduce false alarms in static code analysis tools by combining the BERT deep learning model and a Decision Tree classifier. They used Byte Pair Encoding for word embedding and used a dataset with about 85.9% false-positive percentage, achieving 96% Precision, 92% Accuracy and 94% F1-Score for their BERT-based line-level weakness analysis model. The Alarm Classification model which used DT classifier also achieved 96% F1-Score, 0.89 AUC, and 89% Accuracy. Their proposed models used the Juliet test suite C/C++ 1.3 version as the dataset.

Yerramreddy et al. [8] experimented on three classes of ML techniques, traditional models (such as NBayes and Bayesian networks), LSTM, and Graph Neural Networks, to classify alarms generated by three SCA tools (FindSecBugs, CBMC, and JBMC) into true positives and false-positives. They used similar code pre-processing techniques to Koc et al. [11] and found that LSTM with an "Extracting English Words from Identifiers" transformation produced the best results in terms of Accuracy, Precision and Recall performance metrics, similar to the finding of Koc et al. [11].

Tan and Tians [13] experimented with four ML models to identify false-positive alarms, and Defects4J was used as the dataset, where CNNs produced the best Accuracy and Precision, and BiRNNs had the best Recall. They benchmarked their methods with Lee et al. [10] and Yang et al. [27], finding that their approach was superior when taking average Precision and Recall values into account. Yang et al. [3] explored various CRL techniques to understand the semantics of warnings generated by SCA tools. They used scratch models (Word2Vec, FastText, Glove, and ELMo) and pre-trained models (BERT, RoBERT, and CodeBERT) as CRL techniques. Also, the authors used various ML models (MLP, LSTM, GRU, BiLSTM, BGRU, and TextCNN) for training, finding that CRL models can efficiently differentiate buggy code snippets (median F1-Score of 87.3% for binary classification and 77.4% for multi-class classification) and identify false-positive warnings (F1-Score of best performer being 75.6%). ML models with CRLs such as Word2Vec, FastText, GloVe, and CodeBERT generated higher performance than with other CRLs in binary classifications and when the training dataset contained 36% of bugs. The authors proposed a novel approach called PRISM (**PRI**oritize **S**tatic warnings based on aggregating multiple CRL **M**odels) to prioritise static warnings based on aggregating multiple CRL models to reduce the false-positives generated by existing SCA tools. For CRL models, FastText outperformed other scratch models in terms of both effectiveness and stability (F1-Score) when combined with a Recurrent Neural Network (RNN) CodeBERT and RoBERTa outperformed BERT and performed better in terms of F1-Score when combined with BRNN and TextCNN. For NN models, RNN models perform better than other NN models in terms of F1-Score, but MLP gains performed poorest, showing it may not be ideal for static bug detection.

A summary of the CRLS and ML algorithms that produced high performance in identifying false-positive warnings in relevant existing studies is provided in Table 21 and Table 22, respectively. Here it is noted that various CRL technologies are used for source code prediction studies, including CodeBERT, Word2Vec, Code2Vec, FastText, and GloVe. Among these CRLs, the majority of studies have employed Word2Vec as the CRL. Additionally, CodeBERT and FastText have also been utilised in recent research to represent source code in ML models. To support modelling, various ML algorithms have been employed, including BGRU, GRU, TextCNN, BiLSTM, LSTM, and RF, with LSTM and RF being used most. Given the disconnected manner in which these approaches have been assessed, it is timely to evaluate these technologies in a single study to understand their performance patterns across various datasets.

### *3.1.3 Ranking and prioritising SCA warnings*

Given that some actual warnings may be classified as false-positives when using binary classification, by prioritising warnings based on the likelihood of the warning being true positive, developer may not miss any warning, allowing them to start fixing issues ranked highest. Therefore, ranking algorithms could be useful [24], and are explored in this section.

Ribeiro et al. [28] used an Ensemble Learning method using Adaboost and multiple DTs to rank warnings generated by three SCA tools: Clang Static Analyzer, CPPCheck, and Frama-C. They used only the warning details as the input for model training, where it was shown that the Ensemble Learning method was able to prioritise actual warnings. Ngo et al. [24] introduced

DEFP, a novel approach for ranking static analysis warnings to improve efficiency in identifying actual vulnerabilities. Their approach used Word2Vec for code representation and BiLSTM to train the model. DEFP analyses SCA warnings and prioritises them according to the probability of them being True Positives (TPs). By prioritising warnings that are more probable to represent actual vulnerabilities, DEFP is said to reduce the effort wasted on false-positives, thereby enhancing developer productivity.

Sun et al. [50] proposed an interactive ranking algorithm (IRA) based on developers' vote for warning reports, where IRA performed better than AWARE [51, 52]. The study showed the possibility of using human developers' votes to rank SCA warnings. However, the algorithm used in this study did not use any ML model to rank SCA warnings. Siavvas et al. [53] proposed a ML-based technique for prioritising and classifying security-related static analysis alerts based on alerts' criticality, considering information retrieved from alerts, a vulnerability prediction model, and user feedback. They used DT, NBayes, SVMs, MLP, and RF to classify warnings, finding that MLP provided the best results in terms of Accuracy, Precision and Recall performance metrics. Vu and Vo [23] introduced VULRG, a novel method to rank warnings based on their likelihood of being TPs. VULRG used Word2Vec for word embedding, and CNN and BiGRU were used for model training. They benchmarked VULRG with DEFP and found that VULRG is better than DEFP in terms of Precision and Recall performance metrics in terms of identifying the number of true positive warnings found in top-k% warnings for all k values they have investigated (k= 5, 10, 20, 50, and 60).

With the latest advancement of Large Language Models (LLMs) like BERT, GPT-3, GPT-2, and LLaMA, researchers also classify SCA warnings using LLMs [7]. The best advantage of using those models is that they do not require large dataset for training [7]. Li et al. [7] used ChatGPT to identify false-positives for Use Before Initialization (UBI) Bug suggestions generated by the UBITect bug detection tool [54]. They achieve 80% Accuracy and 94% for soundness and completeness with ChatGPT-4, finding that false-positives can be identified by asking carefully constructed simple questions about function-level behaviours or function summaries from ChatGPT-4. Furthermore, this study showed the ability to classify SCA warnings using LLMs.

### 3.2 Datasets used in existing studies for ML model building

Recent studies have devoted considerable effort to creating warning datasets for ML model development. For instance, Tan and Tian [13] utilised a real bug database known as Defects4J [55] to train several ML models. Yang et al. [3] employed three distinct datasets: a bug dataset developed by Yang et al. [3], Defects4J [55] and a dataset created by Liu et al. [56]. Furthermore, Liu et al. [56] annotated warnings by analysing GitHub commit history. In addition, Yerramreddy et al. [8] used four warning datasets comprising three Java and one C dataset. These datasets were generated using the analysis results of JBMC and CBMC SCA tools. All the datasets utilised by Yerramreddy et al. [8] covered only specific aspects of source code quality, such as vulnerabilities and bugs. Li et al. [7] utilised a UBI (Bug dataset to conduct experiments with ChatGPT, assessing whether bugs were real or false-positive.

Nguyen et al. [6] employed the Juliet test suite (for C/C++) to train ML models. The dataset comprised a collection of test cases organised under 118 distinct CWEs[5]. Nagaraj et al. [5] established a security bug dataset by analysing 2,740 Java files with the FindSecBugs SCA tool. Furthermore, Kharkar et al. [40] compiled a dataset containing 539 null dereference warnings produced by analysing seven Java repositories with the Infer SCA tool, labelled by experienced software developers. Furthermore, Hegedus and Ferenc [14] created a warning dataset by collecting 9,958 GitHub Java projects and analysing them using SonarQube. They analysed commit histories to annotate the dataset. Vu and Vo [23] utilised a dataset that was also employed by Ngo et al. [24]. In total, there were 6,620 warnings across 10 popular open-source C/C++ programs. The dataset consisted of two categories of vulnerability types: Buffer Overflow and Null Pointer Dereference. Additionally, Tanwar et al. [12] created a dataset of 10 CWE categories and labelled them using commit history analysis. Koc et al. [11] utilised two datasets to train their models, the OWASP benchmark and a dataset of warnings generated by the FindSecBugs SCA tool for 14 real-world programs. The first dataset was synthetic, while the second dataset consisted of bugs labelled manually. Lee et al. [10] utilised 9871 alarms, which were either resolved or labelled as false-positives at Samsung Electronics. The dataset comprised warnings generated by their in-house static analysis checkers, covering six types of warnings: Resource Handle Leak, Double Free, Null Pointer Dereference After Null, Tainted Loop Termination Condition, Unintentional Fall Through, and Unreachable Code.

[5] https://samate.nist.gov/SARD/test-suites/112

The datasets used to train ML models were not diverse, as only a limited number of warning types were included [7, 8, 12, 40]. Consequently, none of these datasets covered all aspects of source code quality [57]. Therefore, it is emphasised that building ML models on datasets that encompass various aspects of source code quality would yield generalisable results.

Overall, in this Section we identified that existing studies focus on identifying ML models in terms of performance metrics, such as Precision, Recall and F1-Score. However, existing studies do not completely address the practical purposes of building ML models, as explained in Section 2.2. In our study, we address this gap in addition to evaluating ML models' performance based on a range of factors, including characteristics of training datasets, CRL, and the ML algorithm used to train the model. For our study, we selected the best-performing CRLs: Word2Vec, Code2Vec, CodeBERT, and FastText, as well as the best-performing ML techniques: LSTM, BiLSTM, GRU, BGRU, RF, and TextCNN, to evaluate all the combinations of those approaches (4 * 6 = 24) with four datasets. These datasets have distinct characteristics, and we evaluated all combinations of ML approaches with these datasets to identify performance patterns across different CRLs, ML techniques, and characteristics of datasets. We optimised the hyperparameters of these models and selected the best ones based on their performance on the validation dataset. According to existing studies, identifying the relevant source code slice significantly influences the performance of the ML model [24, 58]. Therefore, we selected the most suitable code-slicing framework from some frameworks used in previous studies for this purpose [11, 24].

## 4 DATASETS

The Stack Overflow dataset collected by Meldrum et al. [57] and the Apache Tomcat source code downloaded from GitHub on 8th December 2023 were analysed using two SCA tools (PMD and SonarQube), making four SCA warning datasets. The original Stack Overflow and Apache Tomcat datasets were selected as these were used in multiple existing studies and are validated as good testbeds [59-64]. Furthermore, PMD and SonarQube were chosen as they were the leading SCA tools within the software engineering practitioner community and were noted for having the lowest false-positive rate [65-67]. Therefore, the four created datasets consist of SCA warnings from code snippets (Stack Overflow) and full source code (Apache Tomcat) for two SCA tools.

A summary of the datasets is presented in Table 3, where warnings were generated from specific rules of SCA tools (Denoted as "# of rules found" in Table 3), with a significantly large number of warnings being generated for each code dataset by each SCA tool. The number of warnings is denoted as "# of alarms" in Table 3. Since manually annotating the whole dataset would demand exorbitant resources, four sample datasets were created randomly [68]. The sample size was calculated with a 95% confidence level and a 5% error rate, and the numbers of samples were divided into each SCA rule proportional to the number of warnings in the population dataset [68, 69]. The number of samples for each dataset is presented as “Sample size” in Table 3. Following the defined standard from our previous study , each SCA warnings in those sample datasets were manually examined to determine whether the warnings were true or false-positives. Moreover, the number of false-positives for each dataset is shown as “# of false-positives in the sample” in Table 3. Furthermore, the number of rules that generated false-positives is denoted as “# of rules generated false-positive” in Table 3 and the false-positive percentage for each dataset is shown as “False-positive percentage” in Table 3. It is noted that certain rules did not produce any false-positive warnings, where all warnings associated with those rules are removed, and the resulting dataset size is referred to as “# of samples for model building” in Table 3. The final false-positive percentages for each dataset are denoted as “False-positive percentage in the final dataset” in Table 3 and these datasets were then used to train and validate the ML models described in this study.

For sample Dataset 1, 531 samples were randomly selected with a 95% confidence level covering all the rules available in the population dataset [71]. Out of these samples, 98 were false-positives. Those false-positives were generated by 13 SCA rules out of 188, the total number of rules that generated warnings. Similarly, sample Dataset 2 was created by selecting 536 samples from Dataset 2, and there were 72 false positive samples from 16 SCA rules out of 197, the overall count of rules that generated warnings. Sample Dataset 3 was created using 646 random samples selected from Dataset 3, and 30 false-positive warnings were found from 5 SCA rules out of 310, the total number of rules that generated warnings. Finally, sample Dataset 4 was created using 631 randomly selected samples from Dataset 4. The sample Dataset 4 consisted of 44 false-positives from 14 SCA rules out of a total number of 281 rules that generated warnings. In order to classify the warnings into true or false positives, 14 defined standards were used. The defined standards are shown in Table 2 with the reason for selecting the standards for our study.

Table 2 Established Standards

| Id | Established Standards | Reason for selecting the standard |
|---|---|---|
| 01 | https://wiki.sei.cmu.edu | Wiki page maintained by Carnegie Mellon University and used as a standard for SonarQube |
| 02 | Book: Clean Code | Used as a standard for SonarQube |
| 03 | Book: Effective Java by Joshua Bloch | Used as a standard for SonarQube |
| 04 | Book: Mastering Regular Expressions by Jeffrey Friedl | Used as a standard for SonarQube |
| 05 | Book: Refactoring: Improving the Design of Existing Code by Martin Fowler, Kent Beck, John Brant, William Opdyke, and Don Roberts | Used as a standard for SonarQube |
| 06 | Google Java Coding Style Guide | Used as a standard for SonarQube and A popular and well-accepted standard [72, 73] |
| 07 | https://cwe.mitre.org/data/definitions/ | Used as a standard for SonarQube and A popular and well-accepted standard [74, 75] |
| 08 | Java Doc | Java official user guide |
| 09 | Book: Object-Oriented Metrics in Practice: Using Software Metrics to Characterize, Evaluate, and Improve the Design of Object-Oriented Systems by Michele Lanza and Radu Marinescu | Used as a reference for PMD |
| 10 | Paper: Linguistic Antipatterns - What They Are and How Developers Perceive Them | [76] |
| 11 | Principle of Least Astonishment | [77] |
| 12 | SOLID principles | Used as a standard for SonarQube |
| 13 | Transitive property of equality | Used as a standard for SonarQube |
| 14 | https://owasp.org | A popular and well-accepted standard [78, 79] |

Table 3 Summary of datasets

| Dataset Name | SCA tool | Code Dataset | # of rules found | # of alarms | Sample size | # of false-positives in the sample | # of rules generated false-positive | False-positive percentage | # of samples for model building | False-positive percentage in the final dataset |
|---|---|---|---|---|---|---|---|---|---|---|
| Dataset 1 | PMD | SOD | 188 | 80788 | 531 | 98 | 13 | 18.45% | 139 | 70.50% |
| Dataset 2 | PMD | AT | 197 | 176293 | 536 | 72 | 16 | 13.43% | 117 | 61.54% |
| Dataset 3 | SonarQube | SOD | 310 | 83890 | 646 | 30 | 5 | 4.64% | 150 | 20.00% |
| Dataset 4 | SonarQube | AT | 281 | 128524 | 631 | 44 | 14 | 6.97% | 298 | 14.77% |

*SOD - Stack Overflow dataset, AT - Apache Tomcat*

The sample datasets were annotated so that false-positive warnings were annotated as “1” and true positive warnings as “0”. Finally, all the rules without any false-positives were removed from the ML model for better learning on the dataset. In Table 3, the number of rules identified by the SCA tools is noted under “# of rules found” and the number of rules that generated false-positive warnings is indicated under “# of rules generated false-positive”. The difference between these figures represents the number of rules removed, as these rules did not produce any false-positive warnings. However, this caused the size of the dataset to be further reduced, as shown in Table 3 under “# of samples for model building”. Otherwise, having only true warnings for most of the SCA rules will be an unnecessary overhead for model performance.

Consequently, the model might start training on the rule names of warnings (e.g. “CommentDefaultAccessModifier”) to predict their label and predict “0” (true) for the SCA warnings belonging to the SCA rules that do not have any false-positive warnings. Thus, warnings from SCA rules that did not have any false-positive samples were removed from the sample sets that were used for model building. On the other hand, as shown in Table 3, all the sample datasets are highly class imbalanced, and to mitigate the class-imbalance problem, several steps were taken, as discussed in the Section 5. Additionally, for simplicity of explanation, in other sections of this paper, all the sample datasets prepared for model building were named Dataset 1, Dataset 2, Dataset 3, and Dataset 4.

## 5 METHODOLOGY

### 5.1 Selecting CRLs and ML algorithms to train models

An informal literature review was conducted using Google Scholar, limited to publications from 2019 to 2024 using the following search string:

("false positive" OR "false positives") source code (learning OR predict OR training OR classify OR classifier OR classification) "static analysis"

As a result of the searches, fifty papers relevant to our study were found by reading their titles and abstracts. These papers were then read entirely, where thirteen relevant papers were selected, which aligns with our study. These studies used CRLs to represent source code and evaluated ML models, in finding false-positive SCA warnings. Referring to the existing papers, four CRLs and six ML algorithms were identified based on their popularity and novelty. These CRLs and ML algorithms are provided in Table 21 and Table 22 respectively. In Table 21 and Table 22, “1” refers to the best approach and “2” refers to the second-best approach, and so on.

As shown in Table 21, most existing papers (8 out of 13) used Word2Vec as a CRL for their studies, so Word2Vec was selected for our study. CodeBERT and FastText were also chosen as most recent studies used them [80]. Code2Vec performs better than Word2Vec for similar tasks [81, 82] and was added to the list. Therefore, Word2Vec, CodeBERT, FastText, and Code2Vec were selected as CRL techniques.

As shown in Table 22, most existing studies used LSTM (3 out of 13) and RF (2 out of 13) to train models to classify static code warnings. However, Yang et al. [3] experimented on several ML algorithms and found that BiLSTM performs better than LSTM. Therefore, both BiLSTM and LSTM are selected for our study. Furthermore, BGRU, TextCNN, and GRU were chosen from recent papers due to their novelty, as these technologies have not been extensively explored in previous studies. Hence, LSTM, BiLSTM, RF, BGRU, GRU, and TextCNN were chosen as the ML algorithms for our study.

Our study evaluates a significant number of different CRLs and ML algorithms. We evaluate CRLs from pretrained models (Code2Vec and CodeBERT) as well as CRLs trained on our datasets (Word2Vec and FastText). Furthermore, we assess CRLs that encompass both static CRLs (Word2Vec, Code2Vec, and FastText) and contextual CRLs (CodeBERT) [3]. Moreover, we examine both simple machine learning algorithms, such as RF and DL models (BiLSTM, LSTM, BGRU, GRU, and TextCNN). Additionally, we assess recently introduced technologies (Code2Vec and BiLSTM) that have not been extensively examined for ML model building to identify false-positive warnings. Moreover, none of the existing studies reviewed thoroughly investigate the impact of various CRLs on ML model performance in detecting false-positive warnings.

In total, we have chosen four CRLs and five ML algorithms to train our four datasets aimed at classifying SCA warnings and identifying false-positive SCA warnings. Furthermore, this study evaluates several ML algorithms and models across different performance metrics, including those optimised for Computational Complexity and minimising false-positives (in RQ2 and RQ3). We assessed four datasets with unique properties to uncover potential patterns in ML model performance related to identifying false-positive warnings (RQ1, RQ2, and RQ3). The training process for building the ML models on each dataset is depicted in Figure 1 and discussed in the Section 5.3.

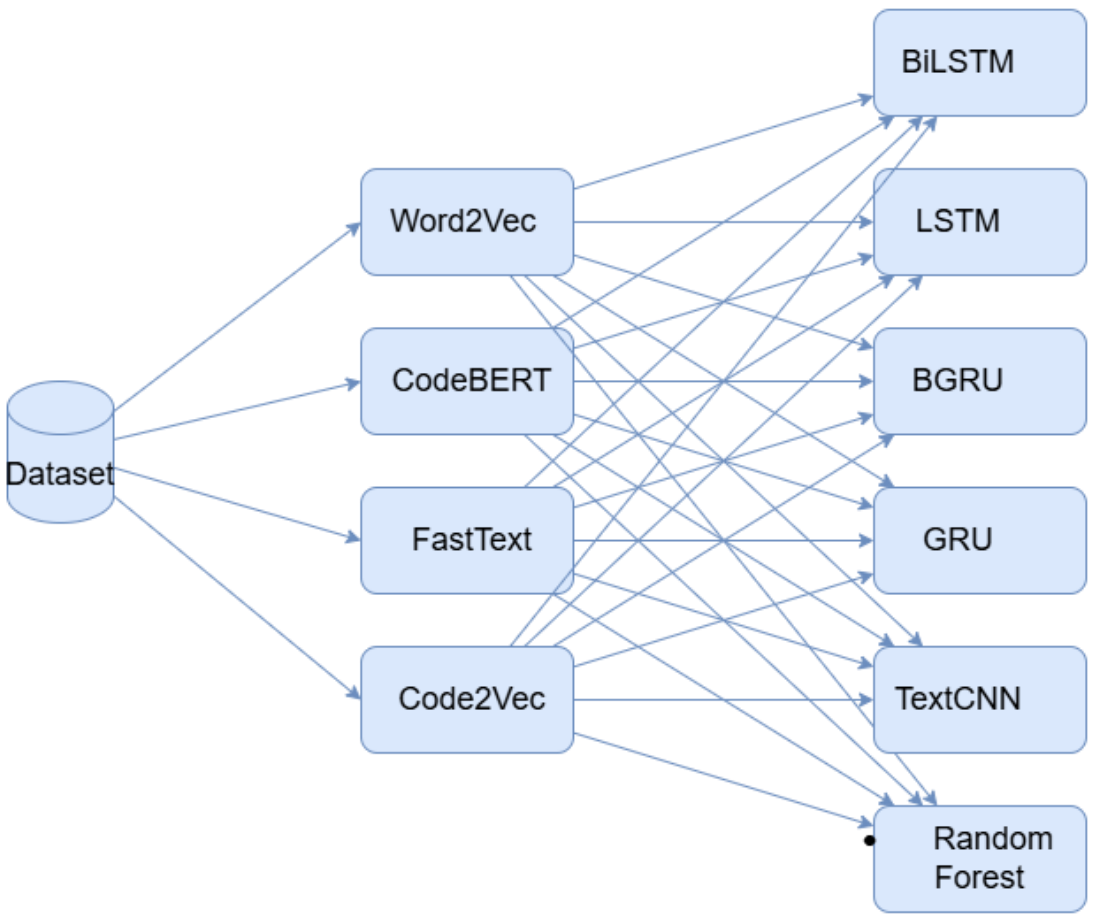

Figure 1 ML model training for a dataset

## 5.2 Data preparation and representation

### *5.2.1 Program slicing*

As shown in Table 4, multiple program slicing approaches were used based on the warning types. While manually annotating the warnings in our previous study, we found two warning types based on how the warning was annotated. In the first type, the warning can be annotated by investigating only a few nearby code lines and does not need to check other blocks of the source code (i.e., the issue is localised to a few nearby code lines) [3], and those types of warnings can be annotated using the issue line or a few code lines near the issue line. In the second type, the issue is globalised to multiple source code blocks. Furthermore, among the lines of source code pertinent to the reported warning, there may be lines of source code that are irrelevant to the issue raised [8]. Thus, for the second type of warning, it is required to check the issue lines and other related issue lines that are far away from the issue line. Sometimes, those related lines were in different scopes of source code.

```java
public class C211053{
Object foo = "x"; // CommentRequired
Integer i = (Integer)foo; //this will

}
```

Type 1 warnings: Only a few lines are needed to determine whether it is a true or a false-positive.

```java
public class LocalXAConnectionFactory implements XAConnectionFactory {

    protected static class LocalXAResource implements XAResource {
        private static final Xid[] EMPTY_XID_ARRAY = {};
        private Xid currentXid; // @GuardedBy("this")

        private Xid checkCurrentXid() throws XAException {
            if (this.currentXid == null) {
                throw new XAException("There is no current transaction");
            }
            return currentXid;
        }

         @Override
         public synchronized void end(final Xid xid, final int flag) throws XAException {
             Objects.requireNonNull(xid, "xid");
             if (!checkCurrentXid().equals(xid)) {
                 throw new XAException("Invalid Xid: expected " + this.currentXid + ", but was " + xid);
             }

         }

    }

}
```

Type 2 warnings: Numerous code lines dispersed throughout the source code are necessary to determine whether it is a true or false-positive.

Figure 2 Code examples for warning types based on how code slices are extracted

One example of the first warning type ("CommentRequired") is shown in Figure 2.A, returned by PMD to identify comments associated only with a specific code line ("inline comments"), which requires only the issue line to be classified. In such a situation, the "ISSUE_LINE_ONLY" approach was used. In contrast, SonarQube Rule Code - "java:S1451" (Track lack of copyright and license headers) requires multiple lines to understand the issue. In such scenarios, either the previous or the following line of code, along with the line where the issue was reported, needs to be extracted. For these cases, either "ONE_LINE_UP" (the issue line and one previous line) or "ONE_LINE_DOWN" (the issue line and one following line) is used.

In the second type of warning, the related issue lines are scattered over the source code file, therefore, extracting the related issue lines is not straightforward for this type of warning. For example, the issue type – "LawOfDemeter" is a second type (as shown in Figure 2.B), where multiple lines of the source code affect the issue. Therefore, following existing studies [8, 83-85], several techniques including "AST_BACKWARD", "AST_FORWARD", "AST_FORWARD_BACKWARD", and "DATA_DEPENDENCY" were used when there is a code flow dependency (Data dependency or AST) when it comes to classifying the second types of warning. Furthermore, "AST_BACKWARD" was used to extract source code lines with AST backward dependencies, while "AST_FORWARD" was used to extract source code lines with AST forward dependencies. If source code lines with both dependencies were required, "AST_FORWARD_BACKWARD" was used. Moreover, "DATA_DEPENDENCY" was used to extract code lines with data dependencies. AST graphs and data dependency graphs are explained in Section 2.3.

Table 4 Source code extraction approaches

| **Program Slicing Approach** | **Description** | **Inspiration** |
|---|---|---|
| ISSUE_LINE_ONLY | Extract the issue line only | [3, 83] |
| ONE_LINE_UP | Extract the issue line and the previous line from the source code | [3, 83] |
| ONE_LINE_DOWN | Extract the issue line and the following line from the source code | [3, 83] |
| AST_BACKWARD | Extract source code lines related to the backward AST graph from the issue line | [8, 83, 84] |
| AST_FORWARD | Extract source code lines related to the forward AST graph from the issue line | [8, 83-85] |
| AST_FORWARD_BACKWARD | Extract source code lines related to the backward AST graph and the forward AST graph from the issue line | [8, 83-85] |
| DATA_DEPENDENCY | Extract source code lines which have data dependencies to the issue line | [8, 83-85] |

Table 4 shows the summary of program slicing approaches used in this study, and the source of their inspiration. The programme slicing approaches used for each PMD warning type are shown in Appendix A, and those used for each SonarQube warning type are shown in Appendix B. A simple script was created to perform the ONE_LINE_UP, ONE_LINE_DOWN, and ISSUE_LINE_ONLY slicing. However, other program slicing types required existing software to identify AST and data dependency graphs (refer to Section 2.3). For such scenarios, existing studies used JOANA[6] and Joern[7] for program slicing [11, 24]. However, since JOANA only supports Java byte code and older versions (JDK 7 and 8) of JVM [86], Joern was used to identify AST and data dependency graphs.

After extracting all the necessary line numbers to classify the warning as either a true positive or false-positive, the corresponding lines of source code were extracted based on the previously identified line numbers. The related warning description was also added to the end of the issue line as a comment (example is shown in Figure 3). This ensured that the CRLs could easily create relationships between the issue and the issue description, because of their proximity [87-89]. The resulting code snippets were used for ML model building.

[6] https://pp.ipd.kit.edu/projects/joana/
[7] https://joern.io/

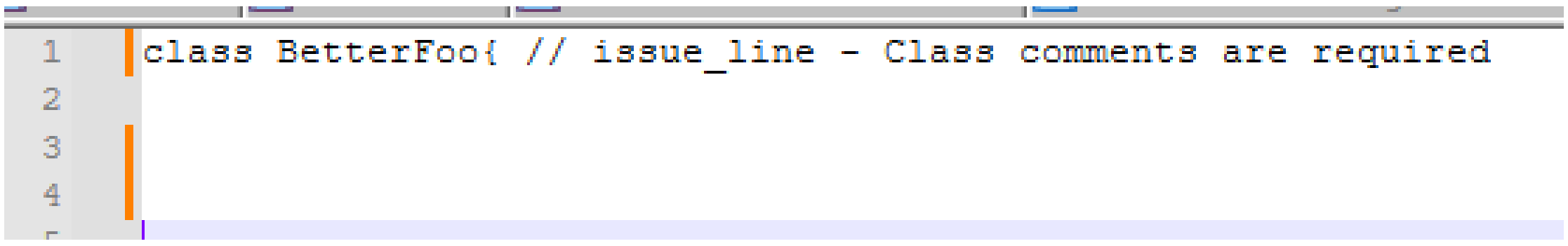

Figure 3 sample code slice with issue description

*5.2.2 Data preparation*

First, relevant code slices were extracted, as explained in Section 5.2.1. However, the prepared data cannot be fed directly to ML algorithms as they expect a numerical vector representation of the dataset [3, 8, 13]. Therefore, data were transformed using the selected CRLs: Word2Vec, CodeBERT, FastText, and Code2Vec. Since pre-trained Word2Vec, and FastText were not trained on source code, our two complete, unprocessed raw source code datasets (Stack Overflow dataset and Apache Tomcat dataset) were used to train these embedding models following Nagaraj et al. [5]. The pre-trained models of CodeBERT and Code2Vec were used in their original form in this study, since those models were trained using Java source code. For CodeBERT, the "microsoft/codebert-base"[8] pre-trained model was used and for Code2Vec, the “Java-large”[9] pre-trained model was used.

The prepared dataset was tokenised, splitting text data into multiple tokens, as CRLs generate vector representations for individual tokens. For example, the code snippet “Object foo = "x"; // issue_line” can be split into multiple tokens such as “Object”, “foo”, “=”, “"”, “x”, “"”, “;”, “//”, “issue_line”. This task was facilitated by the “AutoTokenizer” from the “transformers” Python package, where CodeBERT was used as the source CRL. The Python “pygments” package was used for other RF ML models. In existing studies, the “javalang” Python package is used [3, 8, 14, 90]. However, the “javalang” Python package does not support extracting Java comments, which is required for most scenarios. Therefore, the “pygments” Python package was used. For all the other ML models, “Tokenizer” from “tensorflow.keras.preprocessing.text” Python package was used. The reason behind using multiple tokenizers for different models is that different models require different token formats [13, 14]. For RF models, a flattened vector of word embeddings was needed. However, for other deep learning models, the word embedding vector was created from the embedding matrix in its first (Embedding) layer. CodeBERT has its own tokenizer, and therefore, this default tokenizer was used for all the CodeBERT models.

To use the input vector for the ML models, the input vector size of every code snippet should be standardised to the same length [10, 13, 23, 24]. Therefore, for DL models (models generated by BGRU, GRU, BiLSTM, LSTM, and TextCNN ML algorithms), the input sequence is padded to the length of the largest code snippet size using their tokenizers. Since CodeBERT only supports up to 512-word size, smaller code snippets were padded, and larger code snippets were truncated using its tokenizer. For the other RF models (models with Word2Vec, FastText, and Code2Vec CRLs), the numerical vectors of each code snippet were padded to the size of the largest code snippet vector size by adding zeros at the end [91]. For RandomForest models, except for the CodeBERT-Random_Forest model, this was done after creating vectors using CRLs. Furthermore, padding/truncating was performed first for deep learning models, and then the vector was created using CRLs.

[8] https://github.com/microsoft/CodeBERT
[9] https://github.com/tech-srl/code2vec?tab=readme-ov-file#java-large-compressed-72gb-extracted-37gb

Thereafter, the tokens were vectorised using each CRL with their default parameters. The resulting vector was two-dimensional for the DL models, and considering all data items (all code snippets for all the SCA warnings), it was three-dimensional. In contrast, the input vector was converted to a two-dimensional vector (flattened) for RF, as it does not accept higher dimensional vectors. Then, the resulting vectors, which represent the code slice and the SCA warning description, were fed to the ML algorithm to train the model using each ML algorithm – LSTM, BiLSTM, RF, BGRU, GRU, and TextCNN.

### 5.3 Model training

All four sample datasets were prepared as described in Section 4.3, and the resulting numerical vector was used as the input vector for each ML model. Each dataset was then split into two parts. The first 80% was used for model building, and the remaining 20% was used to test the final model [8, 91]. To perform splitting, "stratify" parameters were enabled to handle the class imbalanced dataset [92].

Hyperparameter optimisation was done using Optuna [93]. The first 80% of the dataset was used to train the models using the Five-Fold cross-validation technique. The stratified version of KFold was employed, as it efficiently split the datasets despite the class-imbalanced nature of the data [92]. In addition, the SMOTE oversampling technique was also used to resolve the class imbalanced problems [94]. SMOTE creates a new dataset using the existing one by oversampling the data points with less frequent labels, ensuring that the resulting dataset does not suffer from class imbalance issues.

When each model was trained, the same dataset needed to be split into training and testing datasets. However, to ensure the data splitting process always generated the same results, a random constant ("42") was used for the random state. The exact number was also used for random seeds in "numpy.random.seed()", "random.seed()", and "tensorflow.random.set_seed()". This ensures consistency of using the same data items for each phase across all ML models.

The early stopping technique, with 'val_loss' as the monitoring value and number 10 as the patience value, was used to train the models [91]. The early stopping technique helps to reduce the chance of overfitting the models [95-99]. Each model trained with 100 epochs and other hyperparameters was optimised using the Optuna [93] with the TPE sampler[10] for 100 iterations for each model. The TPE sampler was the default sampling technique to find the best parameters from the hyperparameter space, and according to Hanifi et al. [100], Optuna with the TPE sampler provides the highest performance. The training dataset was split into five folds [91, 92], and for each of the five iterations, four folds were used to train the model and the other fold was used to validate the model. The validation F1-Score was calculated in each iteration, and the mean value was also calculated. The goal of the optimisation process was to maximise the mean F1-Sscore value. The F1-Score was used for optimisation as the approach is proven to be efficient in minimising the effect of class imbalance problems [101, 102]. The hyperparameters used and the defined ranges for those parameters are shown in Table 5. According to Table 5, the hyperparameters employed for LSTM, BiLSTM, BGRU, and GRU remain consistent. Furthermore, RF and TextCNN possess their own distinct hyperparameters in contrast to other models.

[10] https://optuna.readthedocs.io/en/stable/reference/samplers/generated/optuna.samplers.TPESampler.html#optuna.samplers.TPESampler

Table 5 Hyperparameters with defined ranges

| ML Algorithm | Hyperparameters | Data type | Defined value ranges |
|---|---|---|---|
| LSTM, BiLSTM, BGRU, GRU | n_model_layers | Int | 1 - 3 |
| | learning_rate | Float (Log) | 1e-5 - 1e-2 |
| | batch_size | Int | 4 - 32 |
| | n_dense_layers | Int | 1 – 3 |
| | model_units_l{1… n_model_layers } | Int | 16 - 256 |
| | dense_units_l{1… n_dense_layers} | Int | 16 - 128 |
| | l2_reg_l{1… n_dense_layers} | Float (Log) | 1e-5 - 1e-3 |
| Random Forest | n_estimators | Int | 50 - 300 |
| | max_depth | Int | 3 – 20 |
| | min_samples_split | Int | 2 - 10 |
| | min_samples_leaf | Int | 1 - 10 |
| | max_features | Categorical | 'sqrt' or 'log2' |
| | bootstrap | Categorical | True or False |
| | max_samples (only if bootstrap = True) | Float | 0.5 - 1.0 |
| TextCNN | learning_rate | Float (Log) | 1e-5 - 1e-2 |
| | batch_size | Int | 4 - 32 |
| | n_dense_layers | Int | 1 – 3 |
| | model_l2_reg | Float (Log) | 1e-5 - 1e-3 |
| | n_filters | Int | 16 - 256 |
| | kernel_size_{1 … 3} | Int | 2, 5 |
| | dropout_rate_1 | Float | 0.1 - 0.5 |
| | dropout_rate_1 | Float | 0.1 - 0.5 |
| | dense_units_l{1 … n_dense_layers} | Int | 16 - 128 |
| | dense_l2_reg_l{1 … n_dense_layers} | Float (Log) | 1e-5 - 1e-3 |

Models were trained using Optuna with the hyperparameter ranges outlined Table 5. However, model training requires a considerable amount of time. Therefore, parameter optimisation was conducted on two servers, and the server configurations are presented in Table 6. Additionally, all ML models were evenly distributed across the two servers, as shown in Table 7.

Table 6 Server configurations

| Server ID | Number of CPU cores | RAM (GB) | GPU device |
|---|---|---|---|
| Server 1 | 32 | 128 | GPU A100 80GB |
| Server 2 | 32 | 128 | GPU A100 40GB |

Table 7 Model trained environment

| | Random Forest | GRU | BGRU | LSTM | BiLSTM | TextCNN |
|---|---|---|---|---|---|---|
| *Word2Vec* | Server 1 | Server 0 | Server 1 | Server 0 | Server 1 | Server 0 |
| *Code2Vec* | Server 0 | Server 1 | Server 0 | Server 1 | Server 0 | Server 1 |
| *CodeBERT* | Server 1 | Server 0 | Server 1 | Server 0 | Server 1 | Server 0 |
| *FastText* | Server 0 | Server 1 | Server 0 | Server 1 | Server 0 | Server 1 |

The resulting optimised hyperparameters are shown in Appendix C in our replication package[11]. All the models with the best hyperparameters were retrained on server one sequentially after the initial training process. This ensures consistency across the performance metrics. The same training and testing datasets were used, as was the case in the hyperparameter optimisation phase. While training the ML model, 80% of the training dataset was used to train the model, and 20% was used to validate the model. The validation portion of the training dataset was internally used by the early stopping technique to decide the stopping point [95-99]. Each model was trained 10 times, and the evaluation metrics were collected separately. The final evaluation matrices are shown in Appendix U in our replication package[12] for all datasets. We evaluated the ML models using MCC, F1-Score, Balanced Accuracy and Computational Complexity (Training time). In this context, the Computational Complexity is measured as the time taken to train an ML model, excluding the time needed to generate the numerical vector representation using CRLs. Therefore, the Computational Complexity of CRLs was not included. Additionally, model loss curves for the best trial are shown in Appendix M[13], Appendix O[14], Appendix Q[15], and Appendix S[16] in our replication package, and ROC curves for the best trial are shown in Appendix N[17], Appendix P[18], Appendix R[19], and Appendix T[20] in our replication package. Furthermore, the analysis of the results is discussed in Section 6.

[11] Replication package >> Appendices >> Appendix C_Optimized hyperparameters.docx

### 5.4 Design Of deep learning ML algorithms

The designs of the deep learning models are described in this section. The design of sequential models (GRU [103], BGRU [104], LSTM [105], and BiLSTM [105]) is shown in Figure 4.A and these sequential models were designed with a linear topology (a plain stack of layers). In contrast, TextCNN [91] was designed with non-linear topology as shown in Figure 4.B. Those designs were created following Chollet [91], Gupta et al. [106], Zhang and Wallace [107], Zhang et al. [108], and Zaremba [109].

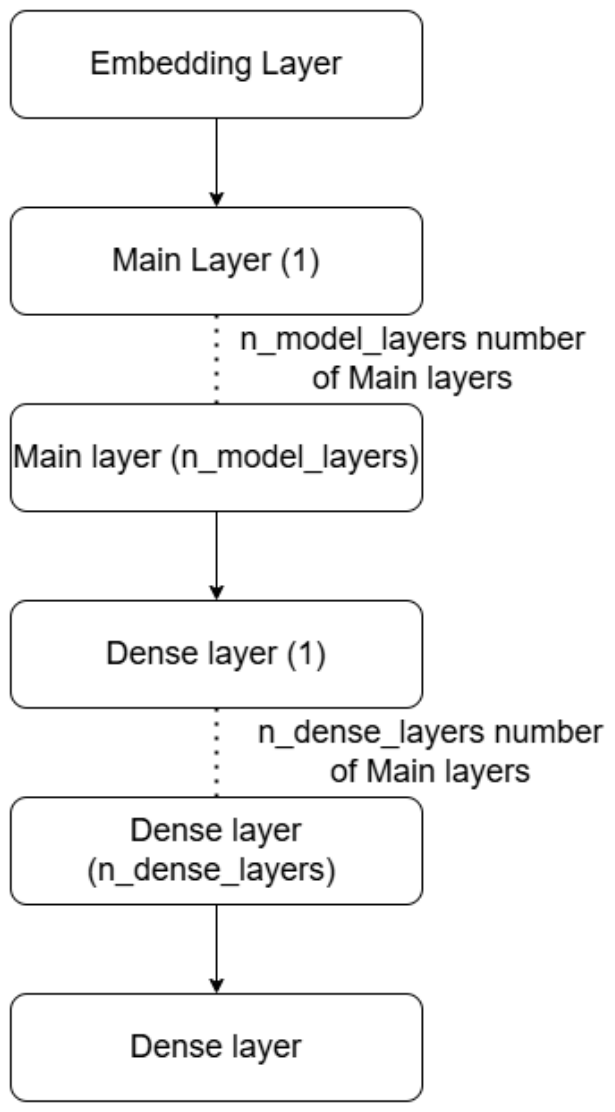

Design of Sequential Models

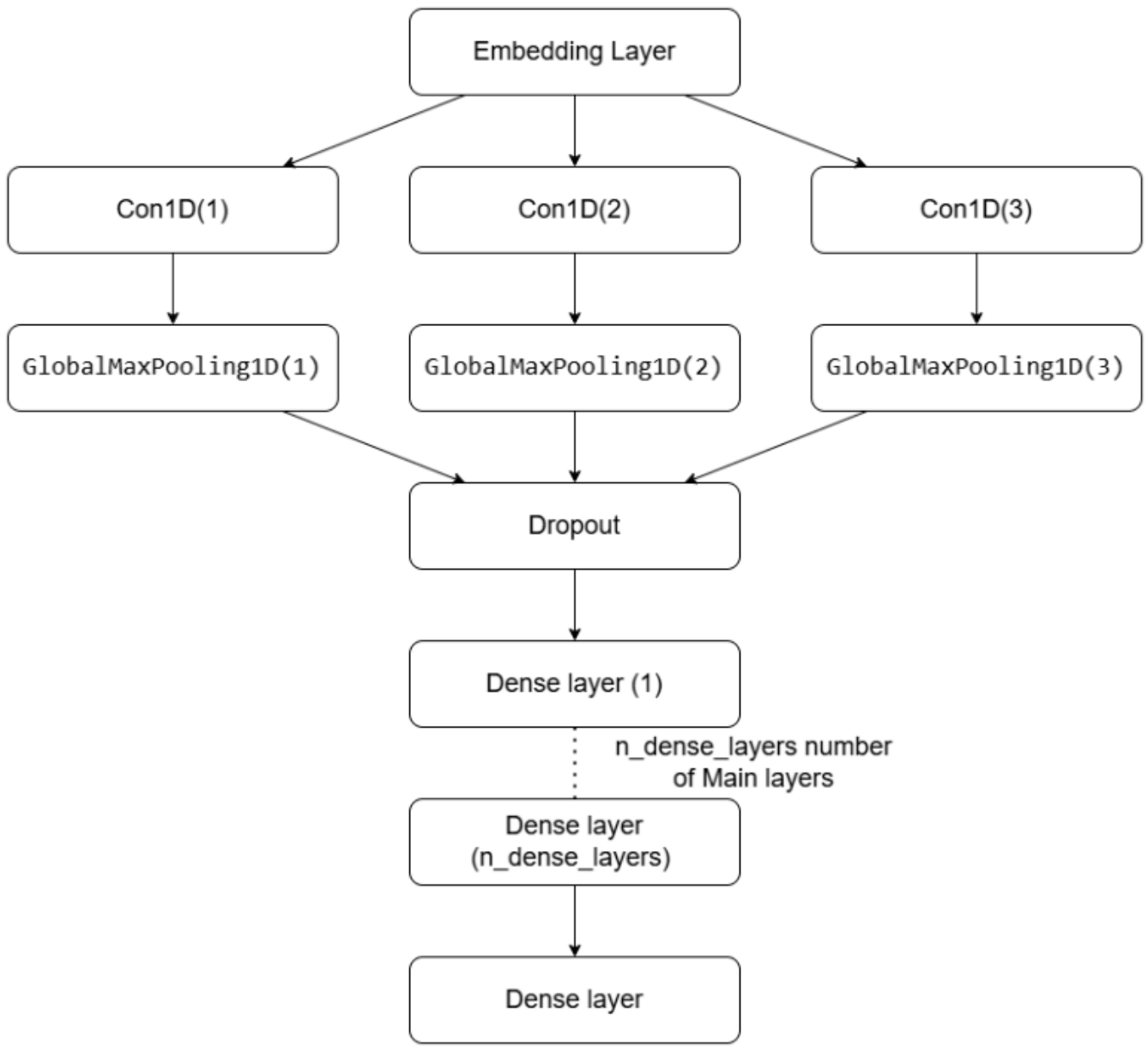

Design of TextCNN Models

[12] Replication package >> Appendices >> Appendix U_Testing results.xlsx
[13] Replication package >> Appendices >> Appendix M_Model Loss for Dataset1.docx
[14] Replication package >> Appendices >> Appendix O_Model Loss for Dataset2.docx
[15] Replication package >> Appendices >> Appendix Q_Model Loss for Dataset3.docx
[16] Replication package >> Appendices >> Appendix S_Model Loss for Dataset4.docx
[17] Replication package >> Appendices >> Appendix N_ROC for Dataset1.docx
[18] Replication package >> Appendices >> Appendix P_ROC for Dataset2.docx
[19] Replication package >> Appendices >> Appendix R_ROC for Dataset3.docx
[20] Replication package >> Appendices >> Appendix T_ROC for Dataset4.docx

Figure 4 Designs of ML Models

As shown in Figure 4.A , for the sequential models, the first layer is the embedding layer, where the word embedding technique is integrated. The following main layers are a variable number (**n_model_layer**) of layers between 1 and 3, decided by the Optuna hyperparameter optimiser. The Main layer is dependent on the model. For example, in GRU models, the Main layer is a GRU layer, while in the BGRU model, the primary layer is a Bi-Directional GRU layer. Likewise, for LSM, it is a LSTM layer, and for BiLSTM, it is a Bi-Directional LSTM layer. The following Dense layers are the variable number (n_dense_layers) layers, where "Relu" is used as the activation function. Finally, there is an output Dense layer, which uses "Softmax" as the activation function.

As shown in Figure 4.B, similar to sequential models, TextCNN also used the Embedding layer to integrate the word embedding technique into the model. Then, there are three parallel one-dimensional convolution layers and a global max pool layer. The three global max-pooling layers are connected to a dropout layer, as shown in Figure 4.B. Later, similar to the sequential model, there are dense layers with similar configurations.

All the DL models used AdamW as the optimiser. It is an improved variant of the Adam Optimiser, which prevents overfitting and allows for a more stable convergence compared to Adam [110]. AdamW was used with "categorical_crossentropy" as the loss function and F1-Score (with 0.5 thresholds) as a metric [111, 112]. The learning rate of the optimiser was selected using hyperparameter optimisation using Optuna, as described in Section 5.3.

## 5.5 Measures for the RQs

We trained the ML models as detailed in the Section 5.3. The four datasets were represented in numerical vector format by source CRLs: Code2Vec, CodeBERT, FastText, and Word2Vec, and those 16 ($4 \times 4$) representations (vectors) were trained using six ML algorithms: BGRU, BiLSTM, GRU, LSTM, RF, and TextCNN (visualised in Figure 1). The resulting 96 ($4 \times 4 \times 6$) ML models were trained while tuning their hyperparameters using Optuna, and their best hyperparameters are shown Appendix C in our replication package. Using the best hyperparameters, models were retrained and tested with the testing dataset 10 times each, producing 960 results. We organised these into a CSV file. The raw data is uploaded to the online repository. Table 8 outlines the data in the CSV file, where each row represents the results of each iteration of the ML model training. Each iteration of the model training involved a code representation technique and an ML algorithm, and each row of the CSV file consists of these details in addition to the dataset ID, iteration number and relevant performance values, as shown in Table 8. Additionally, the training time refers to the duration needed to train the numerical vector representation of the source code and the SCA warning description, excluding the time spent embedding textual inputs such as the source code and the SCA warning.

The methodologies followed in answering each research question are outlined in Table 9. RQ1 evaluates the performance pattern of ML algorithms that vary based on the use of different code representation techniques, while RQ2 examines the performance pattern of ML models that vary based on the use of different ML Algorithms. Moreover, RQ3 assesses that the performance of ML models vary depending on the purpose of building ML models (refer to Section 2.2), while RQ4 examines how the performance of ML models varies with the use of different datasets for each purpose of ML model building. Following these methodologies, we used the Python language to analyse this CSV file and answer the research questions. Additionally, we used statisty.app[21], an online free app to conduct Kruskal-Wallis tests [113-115].

[21] https://statisty.app/kruskal-wallis-test-calculator

The datasets were introduced in Section 4. The focus of this study is to identify false-positive warnings. Therefore, false-positive warnings were annotated as "1", and actual/accurate warnings were annotated as "0". Additionally, the datasets suffer from a class-imbalanced problem, as noted in Section 3, which makes the accuracy metric biased towards the majority class [116, 117]. Therefore, following Tanha et al. [116] and He and Ma [117], MCC, F1-Score, and Balanced Accuracy values were used to evaluate the models' performance. Even though this study focuses on identifying false-positive warnings, reducing misclassifications of actual warnings (True Negatives in our research) can be a practical objective. The Balanced Accuracy values were also calculated to better understand the models' performance for such situations. The usage of these performance metrics for the purpose of model building is discussed in the Section 2.2.

Table 8 Structure of the result CSV file

| Column name | Description |
|---|---|
| Dataset | Dataset ID (e.g. 1) |
| Model | Model name (e.g. – Code2Vec_LSTM) |
| CRL | Code representation technique (e.g. Code2Vec) |
| ML_Algo | ML Algorithm (e.g. - LSTM) |
| Set | Model training iteration number (e.g. 1) |
| MCC | Performance metric |
| F1-Score | Performance metric |
| Balanced Accuracy | Performance metric |
| Training Time | Performance metric |

Table 9 Methodology followed to prepare data to answer RQs

| Research Question | Methodology |
|---|---|
| RQ1 | All four datasets were collectively analysed across different ML algorithms, independent of the datasets used for training and testing the models. For each ML algorithm, the Kruskal-Wallis test, followed by Dunn's post hoc test with a Bonferroni correction, was conducted to evaluate the impact of different CRLs on model performance. The same statistical procedure was applied separately to each performance metric, including MCC, F1-Score, Balanced Accuracy, and Computational Complexity. |
| RQ2 | The four datasets are jointly analysed to evaluate the effect of different ML algorithms on model performance. For each performance metric, the Kruskal-Wallis test is applied, followed by Dunn's post hoc tests with a Bonferroni correction to identify pairwise differences. |
| RQ3 | To determine the best-performing ML models, the results from all datasets were aggregated. For each performance metric, the Kruskal-Wallis test was employed, followed by Dunn's post hoc tests with Bonferroni correction analyses to identify statistically significant pairwise differences between models. |
| RQ4 | The same tests done for RQ3 are conducted for each dataset separately. |

# 6 RESULTS

## 6.1 RQ1: How does the performance of ML algorithms vary with the use of different CRLs in SCA warning classification?

Software source code and SCA warnings were embedded using CRLs to convert the textual inputs into a numerical vector, making the input compatible with ML algorithms. Different CRLs embed textual inputs through various techniques and generate unique numerical vectors. Therefore, this research question examines the impact of utilising various CRLs for each ML algorithm in classifying SCA warnings.

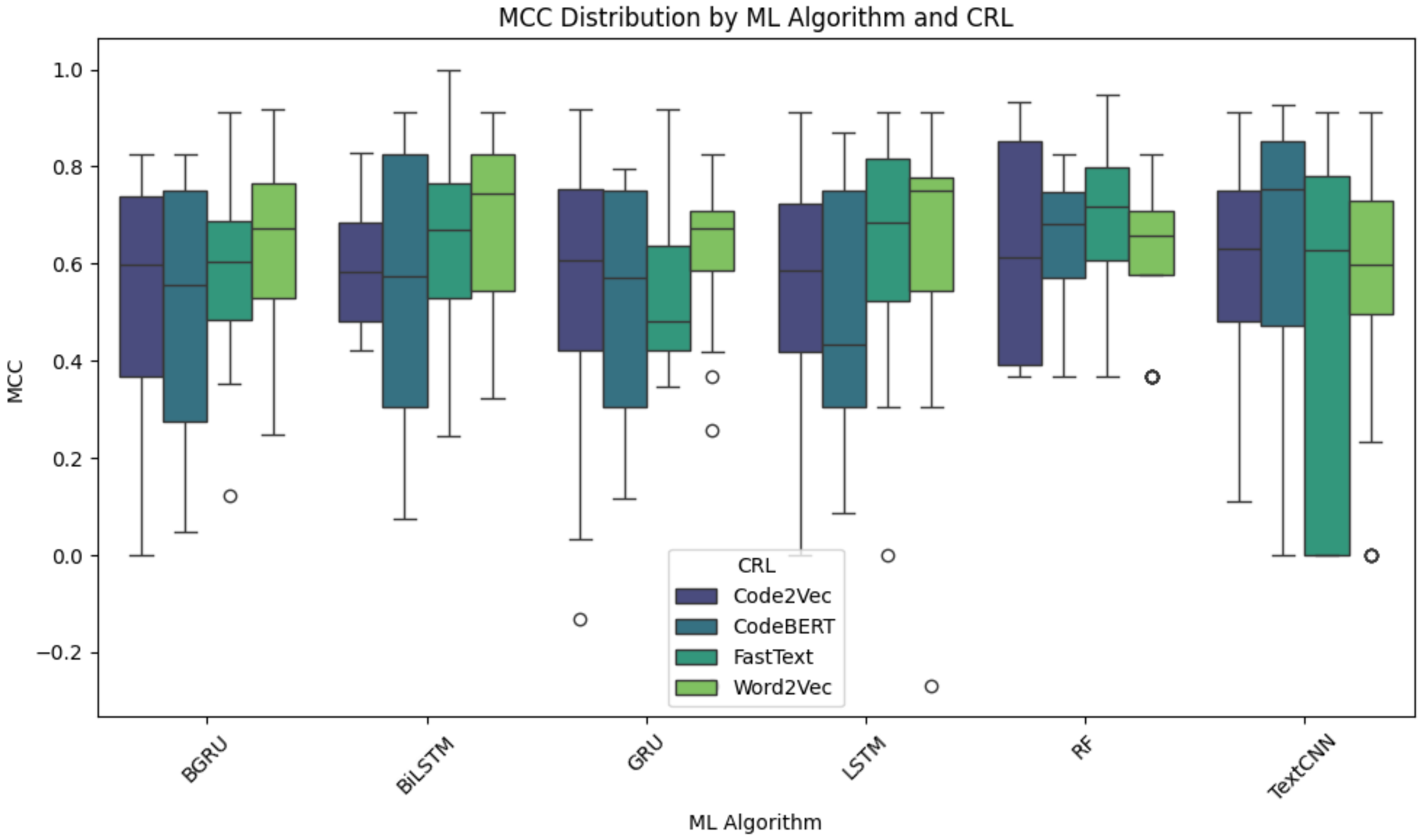

Figure 5 Performance of ML models in each CRL and ML algorithm in MCC

The performance of ML models for each CRL and ML algorithm, measured by the MCC metric, is shown in Figure 5. According to Figure 5, the performance of the ML algorithm varies depending on the CRL used. Word2Vec performs best in BGRU, BiLSTM, GRU, and LSTM, based on median MCC values. However, Word2Vec performs worst in TextCNN, based on median MCC values. The wide boxplots in Figure 5 indicate that the performance of ML models fluctuates randomly when they are trained and tested over time. We conducted Kruskal-Wallis tests between each CRL for all the ML algorithms separately. Table 11 summarises the mean ranks of each CRLs, and the best values are highlighted in bold. Table 11 also shows that Word2Vec performs best in BGRU, BiLSTM, GRU, and LSTM, and worst in TextCNN. Furthermore, according to Table 11, FastText is the best in RF, and CodeBERT is the best in TextCNN. Additionally, the Kruskal-Wallis test results are presented in Table 10, which reveal significant differences only in CRLs for LSTM and TextCNN ($p < 0.05$). Moreover, the Dunn's post hoc tests with a Bonferroni correction comparing CRLs for each ML algorithm are shown in Table 12, indicating that in LSTM, Word2Vec significantly outperforms FastText ($p < 0.05$). Still, there is no significant performance difference between the other combinations. According to Table 12, CodeBERT outperforms both FastText and Word2Vec in TextCNN ($p < 0.05$), and there are no significant performance differences between the other combinations. Thus, RQ1 reveals that **some ML algorithms experience a performance impact depending on the CRL used. However, for most (four out of six) ML algorithms, there are no significant differences in model performance based on the CRL used to generate numerical vectors**.

Similar analyses are performed for F1-Score, Balanced Accuracy, and Computational Complexity performance metrics, and the results are presented in Appendix D in our replication package[22] (from Appendix D.1 to Appendix D.7). According to Appendix D, the best CRL for each ML algorithm in terms of F1-Score and Balanced Accuracy is similar to that of the MCC. However, the performance pattern varies in terms of Computational Complexity, and

[22] Replication package >> Appendices >> Appendix D_Performance analysis for RQ1 in F1-Score, Balance Accuracy and Computational Complexity.docx

CodeBERT surpasses other CRLs when paired with BGRU, BiLSTM, GRU, and LSTM in terms of Computational Complexity. Word2Vec remains the best choice for RF, and Code2Vec is the most suitable for TextCNN in terms of Computational Complexity. There is a significant difference ($p<0.05$) between CRLs in GRU and LSTM for the F1-Score. Furthermore, CRLs in BiLSTM and LSTM are significantly different ($p<0.05$). Conversely, **CRLs across all ML algorithms show significant differences ($p<0.05$) in terms of Computational Complexity**.

Table 10 Kruskal-Wallis Test results between CRLs for each ML algorithm for the MCC performance metric

| **ML Model** | **BGRU** | **BiLSTM** | **GRU** | **LSTM** | **RF** | **TextCNN** |
|---|---|---|---|---|---|---|
| Chi2 | 5.97 | 7.69 | 4.53 | 12.46 | 4.89 | 10.10 |
| df | 3 | 3 | 3 | 3 | 3 | 3 |
| p | 0.113 | 0.053 | 0.210 | 0.006 | 0.180 | 0.018 |

Table 11 Mean rank of CRLs for each ML algorithm for the MCC performance metric (higher is better)

| **CRL** | **BGRU** | **BiLSTM** | **GRU** | **LSTM** | **RF** | **TextCNN** |
|---|---|---|---|---|---|---|
| Code2Vec | 75.80 | 72.40 | 82.35 | 71.13 | 84.25 | 76.28 |
| CodeBERT | 71.86 | 70.20 | 73.83 | 64.78 | 74.25 | 100.46 |
| FastText | 78.99 | 83.76 | 73.35 | 90.10 | 91.75 | 73.01 |
| Word2Vec | 95.35 | 95.64 | 92.48 | 96.00 | 71.75 | 72.25 |

Table 12 Dunn's post hoc test with a Bonferroni correction comparing CRLs for each ML algorithm for the MCC performance metric

| | **Test statistic** | | | | | | **Bonferroni Adjusted Significance** | | | | | |
|---|---|---|---|---|---|---|---|---|---|---|---|---|
| **CRL pairs** | **BGRU** | **BiLSTM** | **GRU** | **LSTM** | **RF** | **TextCNN** | **BGRU** | **BiLSTM** | **GRU** | **LSTM** | **RF** | **TextCNN** |
| Code2Vec - CodeBERT | 3.94 | 2.20 | 8.52 | 6.35 | 10.00 | -24.19 | 1.000 | 1.000 | 1.000 | 1.000 | 1.000 | 0.116 |
| Code2Vec - FastText | -3.19 | -11.36 | 9.00 | -18.97 | -7.50 | 3.26 | 1.000 | 1.000 | 1.000 | 0.401 | 1.000 | 1.000 |
| Code2Vec - Word2Vec | -19.55 | -23.24 | -10.12 | -24.87 | 12.50 | 4.03 | 0.353 | 0.148 | 1.000 | 0.098 | 1.000 | 1.000 |
| CodeBERT - FastText | -7.12 | -13.56 | 0.48 | -25.32 | -17.50 | 27.45 | 1.000 | 1.000 | 1.000 | 0.087 | 0.523 | 0.048 |
| CodeBERT - Word2Vec | -23.49 | -25.44 | -18.65 | -31.22 | 2.50 | 28.21 | 0.139 | 0.084 | 0.429 | 0.015 | 1.000 | 0.038 |
| FastText - Word2Vec | -16.36 | -11.87 | -19.12 | -5.90 | 20.00 | 0.76 | 0.683 | 1.000 | 0.387 | 1.000 | 0.304 | 1.000 |

## 6.2 RQ2: How does SCA warning classification performance vary with the use of different ML algorithms?

ML algorithms such as BGRU, RF, and LSTM are used to develop ML models that identify patterns in training datasets. The way patterns are identified varies depending on the algorithm used. Therefore, it is worthwhile to explore the impact of ML algorithms on SCA warning classification.

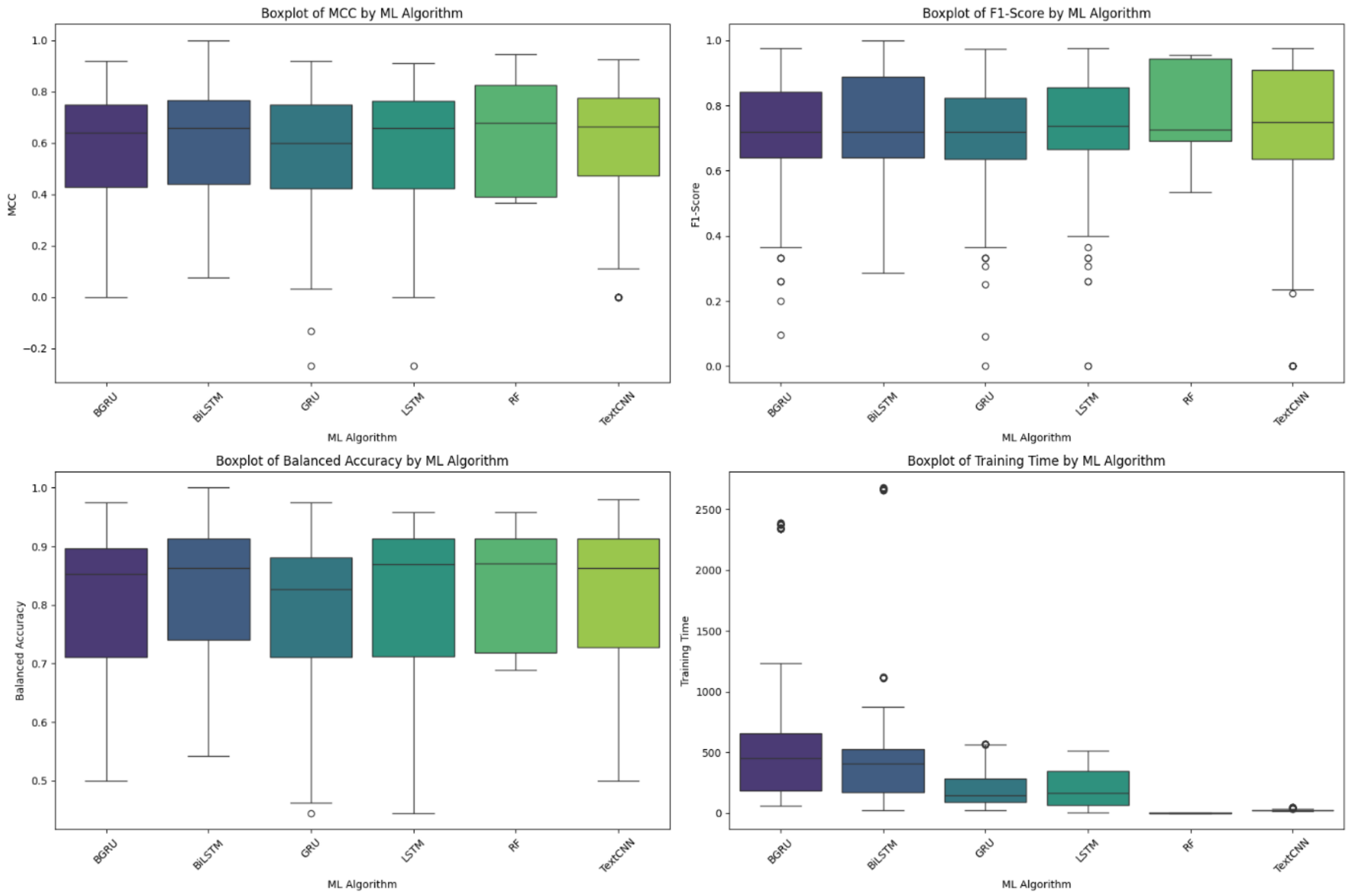

Figure 6 Performance of ML models based on the ML algorithm used

The performance patterns in classifying SCA warnings using MCC, F1-Score, Balanced Accuracy, and Computational Complexity are shown in Figure 6. According to Figure 6, the performance of the model depends on the ML Algorithms used. We conducted Kruskal-Wallis tests between ML algorithms for MCC, F1-Score, Balanced Accuracy and Computational Complexity and the results are shown in Table 13. According to Table 13, there are significant differences ($p<0.05$) in model performance between ML Algorithms in each performance metric. Moreover, Table 14 demonstrates the mean rank of ML algorithms for each performance metric. According to the mean ranks shown in Table 14, Randon Forest is the best in MCC, F1-Score and Computational Complexity. However, BiLSTM is the best in Balanced Accuracy. However, according to the Dunn's post hoc tests with a Bonferroni correction results, which are shown in Table 15, RF significantly outperforms the baseline (GRU) in terms of MCC and F1-Score. Furthermore, BiLSTM, RF, and TextCNN outperform the baseline (GRU) in Balanced Accuracy. Moreover, in Computational Complexity, RF performs significantly better ($p<0.05$) compared to all other ML Algorithms. **Moreover, according to Table 15, most of the ML Algorithms performed significantly differently ($p<0.05$, except BGRU and BiLSTM, and GRU and LSTM) in terms of Computational Complexity)**.

Table 13 Kruskal-Wallis Test results between ML algorithms for each performance metric

| **Metrics** | **MCC** | **F1-Score** | **Balanced Accuracy** | **Training Time** |
|---|---|---|---|---|
| Chi2 | 13.53 | 18.86 | 19.83 | 686.53 |
| df | 5 | 5 | 5 | 5 |
| p | 0.019 | 0.002 | 0.001 | <.001 |

Table 14 Mean rank of ML algorithms for each performance metric (higher is better for MCC, F1-Score and Balanced Accuracy; lower is better for Training Time)

| ML Algorithm | MCC | F1-Score | Balanced Accuracy | Training Time |
|---|---|---|---|---|
| BGRU | 445.55 | 453.52 | 444.15 | 726.59 |
| BiLSTM | 498.81 | 474.22 | 514.48 | 696.70 |
| GRU | 430.74 | 431.06 | 409.08 | 569.91 |
| LSTM | 482.95 | 488.53 | 497.08 | 551.50 |
| RF | **526.75** | **556.41** | 512.56 | **80.66** |
| TextCNN | 498.20 | 479.27 | 505.65 | 257.63 |

Table 15 Dunn's post hoc tests with a Bonferroni correction comparing ML algorithms for each performance metric

| | Test statistic | | | | Bonferroni Adjusted Significance | | | |
|---|---|---|---|---|---|---|---|---|
| **ML Algorithms pairs** | **MCC** | **F1-Score** | **Balanced Accuracy** | **Training Time** | **MCC** | **F1-Score** | **Balanced Accuracy** | **Training Time** |
| BGRU - BiLSTM | -53.26 | -20.71 | -70.34 | 29.88 | 1 | 1 | 0.347 | 1 |
| BGRU - GRU | 14.81 | 22.46 | 35.07 | 156.67 | 1 | 1 | 1 | **<.001** |
| BGRU - LSTM | -37.40 | -35.01 | -52.93 | 175.09 | 1 | 1 | 1 | **<.001** |
| BGRU - RF | -81.20 | -102.89 | -68.42 | 645.92 | 0.131 | **0.013** | 0.408 | **<.001** |
| BGRU - TextCNN | -52.65 | -25.75 | -61.50 | 468.95 | 1 | 1 | 0.706 | **<.001** |
| BiLSTM - GRU | 68.06 | 43.16 | 105.40 | 126.79 | 0.42 | 1 | **0.01** | **0.001** |
| BiLSTM - LSTM | 15.86 | -14.31 | 17.41 | 145.20 | 1 | 1 | 1 | **<.001** |
| BiLSTM - RF | -27.94 | -82.18 | 1.92 | 616.04 | 1 | 0.119 | 1 | **<.001** |
| BiLSTM - TextCNN | 0.61 | -5.05 | 8.83 | 439.07 | 1 | 1 | 1 | **<.001** |
| GRU - LSTM | -52.21 | -57.47 | -87.99 | 18.41 | 1 | 0.952 | 0.067 | 1 |
| GRU - RF | -96.01 | -125.35 | -103.48 | 489.25 | **0.029** | **0.001** | **0.013** | **<.001** |
| GRU - TextCNN | -67.46 | -48.21 | -96.57 | 312.28 | 0.441 | 1 | **0.027** | **<.001** |
| LSTM - RF | -43.80 | -67.88 | -15.49 | 470.84 | 1 | 0.426 | 1 | **<.001** |
| LSTM - TextCNN | -15.25 | 9.26 | -8.57 | 293.87 | 1 | 1 | 1 | **<.001** |
| RF - TextCNN | 28.55 | 77.14 | 6.91 | -176.97 | 1 | 0.191 | 1 | **<.001** |

## 6.3 RQ3: How does SCA warning classification performance vary depending on the purpose of building ML models?

We introduced some potential purposes of ML model building in the context of SCA warning classification in Section 2.2. In this subsection, we examine the performance of ML model variations in relation to the purpose of building ML models. The performance pattern of the ML model reflecting the purpose of the building is shown in Figure 5, Appendix D.1, Appendix D.2 and Appendix D.3. The purpose of the model building is discussed in Section 2.2. To identify the best-performing ML models for each purpose (related to specific performance metrics), Kruskal-Wallis tests were conducted for each performance metric, including MCC,

F1-Score, Balanced Accuracy, and Computational Complexity. According to Table 16, in all performance metrics, at least one model significantly outperforms the other ML models ($p < 0.05$). Furthermore, the mean ranks generated through the analysis shown in Table 17 indicate that CodeBERT_TextCNN is the best in MCC, and FastText_RF is the best in F1-Score. Moreover, according to Table 17, Word2Vec_LSTM is the best in Balanced Accuracy, and Word2Vec_RF is the best in Computational Complexity. Additionally, the Dunn's post hoc tests with a Bonferroni correction are shown in Appendix E in our replication package[23]. According to Appendix E in our replication package, CodeBERT_TextCNN performs significantly better ($p<0.05$) than the baselines (CodeBERT_LSTM, CodeBERT_BGRU, and FastText_GRU) in MCC. Furthermore, in F1-Score, CodeBERT_TextCNN, FastText_RF, Word2Vec_BiLSTM, and Word2Vec_LSTM perform significantly ($p<0.05$) better than FastText_GRU. Moreover, in Balanced Accuracy, Word2Vec_BiLSTM and Word2Vec_LSTM are significantly better than CodeBERT_BGRU. Furthermore, CodeBERT_TextCNN, FastText_BiLSTM, FastText_LSTM, FastText_RF, Word2Vec_BiLSTM, and Word2Vec_LSTM are significantly better than the baseline of both CodeBERT_GRU and CodeBERT_LSTM in terms of Balanced Accuracy. In terms of Computational Complexity, there are no significant differences ($p<0.05$) between CodeBert_LSTM, CodeBERT_TextCNN, FastText_TextCNN, Word2Vec_TextCNN, Code2Vec_TextCNN, CodeBERT_RF, and FastText_RF. Additionally, there are no significant differences ($p < 0.05$) between the RF models. However, Word2Vec_RF, and Code2Vec_RF models are significantly better than CodeBert_LSTM and CodeBERT_TextCNN. Moreover, Word2Vec_RF is slightly better than Code2Vec_RF, according to Table 17. However, there are no significant differences between these models, as noted in Appendix E in our replication package. Consequently, outcomes here suggest that **the optimal ML model varies depending on specific performance metrics, and the best choice of model also changes based on the intended purpose**.

Table 16 Kruskal-Wallis Test results between ML models for each performance metric

| Metrics | MCC | F1-Score | Balanced Accuracy | Training Time |
|---|---|---|---|---|
| Chi2 | 58.44 | 66.01 | 81.91 | 871.93 |
| Df | 23 | 23 | 23 | 23 |
| P | <.001 | <.001 | <.001 | <.001 |

Table 17 Mean rank of ML models for each performance metric (higher is better for MCC, F1-Score and Balanced Accuracy; lower is better for Training Time)

| Groups (n = 40) | MCC | F1-Score | Balanced Accuracy | Training Time |
|---|---|---|---|---|
| Code2Vec_BGRU | 419.45 | 428.60 | 445.30 | 739.05 |
| Code2Vec_BiLSTM | 449.69 | 415.14 | 479.75 | 818.51 |
| Code2Vec_GRU | 442.91 | 429.98 | 457.60 | 552.3 |
| Code2Vec_LSTM | 429.55 | 418.28 | 461.08 | 534.91 |
| Code2Vec_RF | 530.63 | 527.00 | 520.63 | 52.20 |
| Code2Vec_TextCNN | 482.46 | 458.96 | 504.11 | 237.23 |
| CodeBERT_BGRU | 388.86 | 413.05 | 362.45 | 468.80 |
| CodeBERT_BiLSTM | 437.84 | 431.66 | 407.30 | 402.35 |
| CodeBERT_GRU | 394.50 | 415.03 | 322.06 | 393.68 |
| CodeBERT_LSTM | 382.06 | 401.16 | 332.64 | 323.84 |

[23] Replication package >> Appendices >> Appendix E_Dunn's post hoc tests with a Bonferroni correction comparing ML models for each performance metric.docx

| Groups (n = 40) | MCC | F1-Score | Balanced Accuracy | Training Time |
|---|---|---|---|---|
| CodeBERT_RF | 509.13 | 552.13 | 515.50 | 138.85 |
| CodeBERT_TextCNN | 625.30 | 578.11 | 585.45 | 286.14 |
| FastText_BGRU | 441.84 | 442.88 | 444.48 | 878.00 |
| FastText_BiLSTM | 517.42 | 475.70 | 569.29 | 802.95 |
| FastText_GRU | 391.93 | 339.84 | 406.05 | 673.00 |
| FastText_LSTM | 546.58 | 555.65 | 578.45 | 653.83 |
| FastText_RF | 582.00 | 627.63 | 585.50 | 93.13 |
| FastText_TextCNN | 440.48 | 459.00 | 461.81 | 238.36 |
| Word2Vec_BGRU | 532.05 | 529.54 | 524.36 | 820.50 |
| Word2Vec_BiLSTM | 590.28 | 574.39 | 601.60 | 763.00 |
| Word2Vec_GRU | 493.64 | 539.40 | 450.61 | 660.68 |
| Word2Vec_LSTM | 573.61 | 579.03 | 616.14 | 693.43 |
| Word2Vec_RF | 485.25 | 518.88 | 428.63 | 38.48 |
| Word2Vec_TextCNN | 444.56 | 421.00 | 471.23 | 268.81 |

### 6.4 RQ4: How does SCA warning classification performance vary with the use of different datasets?

In answering previous research questions, we analysed the results disregarding the dataset used to train and test ML models. In contrast, this section examines how the choice of dataset influences the performance of the selected machine learning models. The performance patterns for each performance metric and dataset are shown in Appendix F[24], Appendix G[25], Appendix H[26] and Appendix I[27] in our replication package. According to these figures, we can observe distinct performance patterns between the datasets. Therefore, we conducted Kruskal-Wallis tests between datasets for each performance metric. According to the test results shown in Table 18, there are significant differences ($p<0.05$) in performance between each dataset for each performance metric. Therefore, post hoc tests were conducted, and the results are shown in Table 19. The post hoc test results (Table 19) indicate that there are significant differences in performance between each dataset in all the performance metrics.

According to Table 19, there are no significant differences between Dataset 1 and Dataset 2, nor between Dataset 3 and Dataset 4 in Computational Complexity. Also, there are no significant difference between Dataset 2 and Dataset 3 in terms of F1-Score. However, there are substantial performance differences ($p<0.05$) between other combinations of dataset pairs in each performance metrics. To identify ML model performance patterns within each dataset, we then conducted Kruskal-Wallis tests comparing each ML model separately across performance metrics and datasets, and the results are shown in Appendix J in our replication package[28]. The mean rankings for ML models are presented in Appendix K in our replication package[29] (for Dataset 1 and Dataset 2) and Appendix L in our replication package[30] (for Dataset 3 and Dataset 4). The best mean ranking values are highlighted in bold. According to Appendices M and N, **the best-performing ML model depends on the dataset used**. Moreover, according to Appendix K in our replication package and Appendix L in our replication package, **the best ML model for each dataset and performance metric varies**, and the best ML models for different scenarios are shown in Table 20. According to Table 20, the best-performing ML model varies based on the datasets. Moreover, RF models with either Word2Vec or Code2Vec CRLs are best in terms of Computational Complexity for all the datasets. Additionally, FastText CRL with ML algorithms such as TextCNN, BiLSTM, BGRU, and LSTM are the best in Dataset 1, Dataset 2, and Dataset 3 in terms of MCC, F1-Score, and Balanced Accuracy. For Dataset 4, Code2Vec_RF performs the best across all performance metrics. Outcomes here suggest that **the performance of SCA warning classification depends on the characteristics of the dataset used to train and test ML models**.

[24] Replication package >> Appendices >> Appendix F_Performance of ML models in each dataset in MCC.docx
[25] Replication package >> Appendices >> Appendix G_Performance of ML models in each dataset in F1-Score.docx
[26] Replication package >> Appendices >> Appendix H_Performance of ML models in each dataset in Balanced Accuracy.docx
[27] Replication package >> Appendices >> Appendix I_Performance of ML models in each dataset in Computational Complexity.docx
[28] Replication package >> Appendices >> Appendix J_Kruskal-Wallis Test results between ML models separately in different Datasets and Performance metrics.docx
[29] Replication package >> Appendices >> Appendix K_Mean rank of ML models separately in different Datasets (Dataset 1 and Dataset 2) and Performance metrics.docx
[30] Replication package >> Appendices >> Appendix L_Mean rank of ML models separately in different Datasets (Dataset 3 and Dataset 4) and Performance metrics.docx

Table 18 Kruskal-Wallis Test results between datasets

| **Metrics** | **MCC** | **F1-Score** | **Balanced Accuracy** | **Training Time** |
|---|---|---|---|---|
| Chi2 | 518.47 | 510.76 | 455.92 | 21.98 |
| df | 3 | 3 | 3 | 3 |
| p | <.001 | <.001 | <.001 | <.001 |

Table 19 Dunn's post hoc tests with a Bonferroni correction comparing performance of datasets

| | **Test statistic** | | | | **Bonferroni Adjusted Significance** | | | |
|---|---|---|---|---|---|---|---|---|
| **Dataset pair** | **MCC** | **F1-Score** | **Balanced Accuracy** | **Training Time** | **MCC** | **F1-Score** | **Balanced Accuracy** | **Training Time** |
| 1 - 2 | 550.56 | 503.27 | 506.43 | -13.33 | **<.001** | **<.001** | **<.001** | 1 |
| 1 - 3 | 333.87 | 485.84 | 231.95 | -90.10 | **<.001** | **<.001** | **<.001** | **0.002** |
| 1 - 4 | 168.53 | 335.26 | 94.40 | -89.99 | **<.001** | **<.001** | **0.001** | **0.002** |
| 2 - 3 | -216.69 | -17.43 | -274.48 | -76.77 | **<.001** | 1 | **<.001** | **0.015** |
| 2 - 4 | -382.03 | -168.01 | -412.03 | -76.66 | **<.001** | **<.001** | **<.001** | **0.015** |
| 3 - 4 | -165.34 | -150.58 | -137.55 | 0.11 | **<.001** | **<.001** | **<.001** | 1 |

Table 20 Best ML models for each dataset and performance metric

| **Datasets** | **MCC** | **F1-Score** | **Balanced Accuracy** | **Training Time** |
|---|---|---|---|---|
| Dataset 1 | FastText_TextCNN | FastText_TextCNN | FastText_BiLSTM | Word2Vec_RF |
| Dataset 2 | FastText_BGRU | FastText_LSTM | FastText_BGRU | Word2Vec_RF |
| Dataset 3 | FastText_RF | FastText_RF | FastText_RF | Code2Vec_RF |
| Dataset 4 | Code2Vec_RF | Code2Vec_RF | Code2Vec_RF | Code2Vec_RF |

## 7 DISCUSSION AND IMPLICATIONS

Existing studies have shown that SCA tools generate a significant amount of false-positive warnings, which wastes developers' time, as it takes considerable effort to identify actual warnings that need to be fixed [1, 2]. Therefore, several efforts have already been invested to identify false-positive warnings, either by classifying SCA warnings or by ranking SCA warnings [3, 6, 8, 13, 50, 52, 53]. Moreover, there are a few attempts using LLMs for identifying false-positive SCA warnings [7]. However, these existing studies do not entirely address practical problems associated with classifying SCA warnings, such as understanding the potential purposes of building ML models to classify SCA warnings. Instead, existing studies have tried to optimise ML models with a few performance metrics without focusing on these purposes. Therefore, we identified four primary goals in Section 2.2 which are applicable to classify SCA warnings and now discuss our research questions below.

### 7.1 RQ1: How does the performance of ML algorithms vary with the use of different CRLs in SCA warning classification?

For RQ1, all the datasets were analysed together across different ML algorithms, including Word2Vec, Code2Vec, FastText, and CodeBERT, independent of the datasets used for the analysis. For each ML algorithm, the Kruskal-Wallis test followed by Dunn's post hoc test with a Bonferroni correction was performed to evaluate model performance across different CRLs (Word2Vec, Code2Vec, FastText, and CodeBERT). The same statistical analysis was conducted separately for each performance metric.

Findings from RQ1 show that the variant of CRL affects the performance of the ML algorithm, where the optimal CRL varies between different ML algorithms. For MCC, F1-Score, and Balanced Accuracy, Word2Vec excels with BGRU, BiLSTM, GRU, and LSTM. Additionally, FastText is the most effective for RF, and CodeBERT is the most effective for TextCNN in terms of MCC, F1-Score, and Balanced Accuracy. Furthermore, in terms of Computational Complexity, CodeBERT outperforms other CRLs in BGRU, BiLSTM, GRU, and LSTM. For Computational Complexity, Word2Vec is best suited for RF, while Code2Vec is optimal for TextCNN.

We found that there were significant performance differences between each CRL only for a few ML algorithms, indicating that we can use any CRL with other ML algorithms without compromising substantial performance. For example, only LSTM and TextCNN showed significant performance differences in the MCC metric between CRLs, indicating that any CRL would perform similarly to BGRU, BiLSTM, GRU, and RF without any notable differences in MCC. Conversely, for optimal model performance in MCC, Word2Vec should be combined with LSTM, and CodeBERT should be paired with TextCNN.

These findings indicate that the effectiveness of a CRL varies considerably across different performance metrics depending on the underlying ML algorithm, rather than being consistent. Word2Vec's strong performance with RNN-based models demonstrates that dense token-level embeddings are effective for sequential modelling approaches [103, 118-120]. Conversely, FastText embeddings seem to complement tree-based models, such as RF, likely because of the additional subword information that enhances feature diversity [121, 122]. Similarly, CodeBERT's pretrained representations of both natural language and source code benefit CNN-based models, which are adept at extracting semantic features from dense vectors [123, 124]. Conversely, Code2Vec consistently underperformed across all ML algorithms in this study. This is likely because Code2Vec is designed to learn from AST paths, which capture structural information but may not align well with the token-sequence–based classifiers used here [81, 103, 120]. Therefore, these results highlight that the interaction between the algorithm and CRL must be carefully considered, as not all code-specific representations automatically lead to improved classification performance.

Practically, these findings imply that the practitioner community may prefer lighter CRLs (in terms of training CRLs with their own source code), such as Word2Vec or FastText, when using traditional or recurrent models, reserving CodeBERT for CNN-based architectures despite its increased computational cost (considering the computational cost involved in embedding source code and SCA warning description with CodeBERT).

### 7.2 RQ2: How does SCA warning classification performance vary with the use of different ML algorithms?

For RQ2, all datasets are analysed collectively to evaluate the impact of different ML algorithms on model performance. For each performance metric, the Kruskal-Wallis test is conducted, followed by Dunn's post hoc tests with a Bonferroni correction to identify pairwise differences. The results for RQ2 show that the performance of SCA warning classification varies according to both the ML algorithm and the evaluation metric. RF achieves the best results in terms of MCC and F1-Score while also requiring the lowest Computational Complexity. BiLSTM, however, perform best in terms of Balanced Accuracy, although the performance difference between BiLSTM and RF in this metric was not statistically significant. Therefore, RF may be considered the most suitable algorithm for general-purpose model building. Overall, the findings indicate that selecting an ML algorithm significantly influences the performance of SCA warning classification. In contrast, results from RQ2 demonstrate that the choice of evaluation metric has a comparatively minor effect.

The better performance of RF in terms of MCC and F1-Score can be attributed to the ensemble nature of the algorithm, which effectively reduces overfitting and performs reliably in cases of class imbalance [122, 125]. Its lower Computational Complexity also highlights the efficiency of traditional ML approaches in settings with limited data. In contrast, BiLSTM's advantage in Balanced Accuracy highlights the strength of bidirectional sequence modelling in capturing contextual dependencies, which seems particularly beneficial for warnings in the minority class [104, 105, 126]. The underperformance of GRU and LSTM compared to BGRU and BiLSTM suggests that unidirectional models are less effective at modelling the bidirectional dependencies inherent in source code [104, 105, 126]. Similarly, the relatively good results of TextCNN indicate that it is sufficient to extract local features in most warnings to capture the sequential and semantic characteristics of SCA warnings and source code [124]. Additionally, it has benefited from advanced program slicing techniques and appending SCA warning messages near the issue line.

Overall, these findings demonstrate that the choice of algorithm has a significant influence on classification outcomes, with RF generally outperforming other algorithms. Consequently, the results suggest that the practitioner community may prefer to use RF, as it provides strong performance with lower Computational Complexity.

### 7.3 RQ3: How does SCA warning classification performance vary depending on the purpose of building ML models?

For RQ3, we determine the best-performing ML models, and for this purpose, the results from all datasets were aggregated. For each performance metric, the Kruskal-Wallis test was employed, followed by Dunn's post hoc tests with a Bonferroni correction analyses to identify statistically significant pairwise differences between models. Outcomes from RQ3 reveal significant differences between ML models (CRLs paired with ML algorithms) within each performance metric. According to the findings, CodeBERT_TextCNN performs best in terms of MCC, while FastText_RF is the top performer for F1-Score. Additionally, Word2Vec_LSTM performs best in terms of Balanced Accuracy, while Word2Vec_RF excels in terms of Computational Complexity. Comparing the results from RQ1 and RQ2, the results generated for RQ3 differ slightly. According to RQ1 and RQ2, FastText_RF was the best ML model in terms of MCC, F1-Score, and Balanced Accuracy. However, in RQ3, FastText_RF is the best only in F1-Score. Considering all these results show that model performance depends on the variant of both the CRL and ML algorithms used to build the model.

CodeBERT captures rich semantic information from source code, while TextCNN excels at detecting local patterns [123, 124]. Therefore, CodeBERT_TextCNN may achieve the best overall performance in SCA warning classification. FastText represents words (or tokens) using subword information, enabling it to generate useful embeddings even for rare or unseen tokens, such as uncommon variable names in code [121]. Moreover, RF is an ensemble of DTs, with each tree learning a different perspective of the data; the forest then takes a majority vote [122]. This improves its ability to recognise a broader range of source code patterns and effectively identify nonlinear relationships, making RF suitable for detecting false-positive warnings. As a result, the FastText_RF combination generally achieves a good balance between Precision and Recall, thereby boosting the F1-Score.

Word2Vec maps tokens into dense vectors that retain semantic similarity, grouping warnings with similar wording (e.g., "unused variable" vs. "redundant variable") together in the vector space [118, 127]. This provides the model with a compact yet meaningful representation of SCA warnings [118, 127]. Additionally, LSTM is designed to learn dependencies in sequences, making it more effective at tracking sequential information than ML algorithms like RF [105, 119, 120, 122]. This may help capture minority classes more accurately, improving the handling of class-imbalanced datasets and increasing Balanced Accuracy. Since both Word2Vec and RF are lightweight, with low Computational Complexity, Word2Vec_RF could achieve the best performance in terms of computational efficiency. Therefore, CodeBERT_TextCNN can be used if overall performance is the priority, FastText_RF for detecting false-positive warnings, and Word2Vec_LSTM for identifying actual warnings with minimal false-positives. Word2Vec_RF is also suitable for faster training and evaluation of ML models with limited computational resources, allowing for fine-tuning to improve performance.

### 7.4 RQ4: How does SCA warning classification performance vary with the use of different datasets?

Each dataset possesses unique characteristics, as outlined in Section 4. Datasets 1 and 2 (PMD) have substantially higher false-positive percentages (70.5% and 61.54%), whereas Datasets 3 and 4 (SonarQube) have much lower false-positive percentages (20.00% and 14.77%). In terms of source code length, Datasets 1 and 3 were derived from code snippets, while Datasets 2 and 4 come from the Apache Tomcat full-length source code. Consequently, the unique characteristics of the datasets motivated RQ4 to examine performance variability across ML models and datasets. Therefore, for each dataset separately, the Kruskal-Wallis test was employed to compare each ML model, followed by Dunn's post hoc tests with a Bonferroni correction to identify statistically significant pairwise differences between models. Kruskal-Wallis tests indicate significant differences in performance across datasets.

The experimental results demonstrate a clear relationship between dataset characteristics and the performance of the ML models. Datasets 1 and 2, which were generated using PMD and contain a high percentage of false-positive warnings, required more sophisticated deep learning architectures to achieve satisfactory performance. For example, the best-performing models across most metrics for Dataset 1 were FastText-based deep learning variants (TextCNN and BiLSTM), while Dataset 2 also favoured FastText-based recurrent architectures (BGRU and LSTM). The majority of the warnings in Datasets 1 and 2 consist of simple warnings which require only investigating a single line of source code. Related rules for these warnings include CommentRequired, MethodArgumentCouldBeFinal and NoPackage. Therefore, FastText and Word2Vec are efficient in representing such simple source codes [121]. Additionally, FastText and Word2Vec are trained with our source code datasets, which boosts their efficiency in representing source codes related to warnings in Datasets 1 and 2. Further, because of a higher false-positive percentage, there are more variations of patterns in the SCA warnings in Datasets 1 and 2, even though source codes related to these warnings are simple. Therefore, these datasets have benefited from the complex pattern identification capability of DL algorithms [103-105, 119, 120, 126].

In contrast, Datasets 3 and 4, generated by SonarQube with significantly lower false-positive percentages, exhibited different trends. For Dataset 3, the classical ML algorithm (RF) with FastText embeddings (FastText_RF) consistently outperformed DL variants in MCC, F1-Score, and Balanced Accuracy, suggesting that when warnings have fewer false-positives, simpler models may suffice. Similarly, the Code2Vec_RF model outperformed on Dataset 4 across all performance metrics considered in this study, further supporting the fact that, with a lower false-positive percentage, classical ML algorithms such as RF perform best. Furthermore, in both Datasets 3 and 4, the majority of the warnings were generated for the SonarQube rule "Source code should be indented consistently" and "Magic numbers should not be used". However, their proportions (number of warnings for the rule "Source code should be indented consistently" divided by the number of warnings for the rule "Magic numbers should not be used") differ. For Dataset 3, this proportion is 2.5, while in Dataset 4, it is 12. Additionally, in Dataset 4, there are warnings for more variety of rules. Therefore, Dataset 3 can be represented with simpler word embedding models such as FastText, while Dataset 4 requires complex CRL models such as Code2Vec [81].

Another key observation relates to training efficiency. For both Dataset 1 and Dataset 2, Word2Vec with RF achieved the best Computational Complexity. In contrast, for Datasets 3 and 4, Code2Vec with RF achieve the best Computational Complexity. Out of the six ML algorithms used in this study, RF is the fastest ML algorithm for training ML models. Meanwhile, Word2Vec is the fastest and simplest among the four CRLs used in the study. Thus, RF is the best ML Algorithm for every dataset in terms of Computational Complexity. Datasets 1 and 2 may have many SCA warnings, which only requires a single issue line to classify SCA warnings. Therefore, for these datasets, Word2Vec provides the best performance in terms of Computational Complexity. However, Datasets 3 and 4 might require more complex and semantic source code representations. As a result, Code2Vec may produce more compatible code representations than Word2Vec and FastText for Datasets 3 and 4. Therefore, Code2Vec_RF provided the best performance for Datasets 3 and 4 in terms of Computational Complexity. Moreover, comparing dataset characteristics with classification results shows that source code length does not significantly impact performance, likely due to the use of improved code-slicing techniques that extract only the relevant portions of the source code. Consequently, higher false-positive percentages appear to necessitate more sophisticated deep learning architectures to compensate for noise. In comparison, datasets with lower false-positive percentages can be effectively modelled with simpler approaches.

Table 21 CRL techniques used in recent papers

| Paper | Published Year | CodeBERT | Word2Vec | Code2Vec | CodeBert CPP | FastText |
|---|---|---|---|---|---|---|
| [13] | 2024 | 1 | | | | |
| [3] | 2024 | | 1 | | | 1 |
| [8] | 2023 | | 1 | | | |
| [7] | 2023 | | | | | |
| [6] | 2023 | | | | 1 | |
| [5] | 2022 | | 1 | | | |
| [40] | 2022 | | | | | |
| [14] | 2022 | | 1 | | | |
| [23] | 2022 | | 1 | | | |
| [12] | 2021 | | | 1 | | |
| [24] | 2021 | | 1 | | | |
| [11] | 2019 | | 1 | | | |
| [10] | 2019 | | 1 | | | |

Table 22 Machine Learning techniques used in recent papers

| **Paper ID** | **1** | **2** | **3** | **4** | **5** | **6** | **7** | **8** | **9** | **10** | **11** | **12** | **13** |
|---|---|---|---|---|---|---|---|---|---|---|---|---|---|
| Paper Title | [13] | [3] | [8] | [7] | [6] | [5] | [40] | [14] | [23] | [12] | [24] | [11] | [10] |
| Published Year | 2024 | 2024 | 2023 | 2023 | 2023 | 2022 | 2022 | 2022 | 2022 | 2021 | 2021 | 2019 | 2019 |
| ChatGPT | | | | 1 | | | | | | | | | |
| BGRU | 1 | 1 | | | | | | | | | | | |
| GRU | | 1 | | | | | | | | | | | |
| TextCNN | | 1 | | | | | | | | | | | |
| DT model | | | | | | | | 2 | | | | | |
| LSTM-ext | | | 1 | | | | | | | | | 1 | |
| LSTM | | 2 | | | | | | | | 1 | | | |
| GPT-C based on GPT-2 | | | | | | | 1 | | | | | | |
| DeepInferEnhance based on CodeBERTa2 | | | | | | | 2 | | | | | | |
| Random forest | | | | | | 1 | | 1 | | | | | |
| CNNs | 2 | | | | | | | | 3 | | 2 | | 1 |
| BiLSTM | | 1 | | | | | | | 2 | | 1 | | |
| CNN+BiGRU | | | | | | | | | 1 | | | | |
| BERT | | | | | 1 | | | | | | | | |
| Ensemble ( XGBoost + Random Forest + Neural Network) | | | | | | | | | | 1 | | | |
| Graph Neural Networks | | | 2 | | | | | | | | | 2 | |
| MLP | 3 | 2 | | | | | | | | | | | |
| Attention mechanism | 3 | | | | | | | | | | | | |
| XGBoost | | | | | | 3 | | | | | | | |
| Ensemble ( Random forest + XGBoost + SVM) | | | | | | 2 | | | | | | | |
| SVM | | | | | | 4 | | | | | | | |
| NBayes | | | | | | | | 2 | | | | | |
| NeuralNet | | | | | | | | 2 | | | | | |

The best CRLs (or the only CRL used for their study) in each existing study are shown in Table 21 and the best ML algorithms in each existing study are shown in Table 22. According to Table 21, Word2Vec is the best in many existing studies. However, according to Table 22, we cannot identify any best ML algorithm from existing studies. Considering the existing studies discussed in the Section 3, none of the existing studies have comprehensively analysed many ML models with combinations of various CRLs and ML Algorithms, in terms of various perspectives, such as multiple performance metrics, model-building goals, and multiple datasets with unique characteristics. Therefore, we have selected four CRLs and six ML algorithms used in existing studies and evaluated these using four SCA warning datasets. Results generated from our study show that the best CRL and ML algorithm is based on the characteristics of the dataset and the purpose of the model building. Nagaraj et al. [5] identified that even simple models can produce significant performance, and our study yields the same results as those of Nagaraj et al. [5]. Moreover, some studies only used specific warning types [5-8]. However, in our study, we used all the warning types which generated false-positive warnings for our two source code datasets by PMD and SonarQube. Furthermore, existing studies optimised ML models without considering the purpose of model building [10-12, 21]. The majority of them labelled false-positives of SCA warnings as “1” and “0” for actual warnings, and they optimised their models with F1-Score. However, as explained in Section 2.2, these models are optimised to find false-positive warnings, but they do not completely cater to all the identified purposes. To address this gap, we identified four possible purposes of model building and identified the best performance metrics to achieve these purposes. We then analysed the performance of ML models based on these performance metrics to find the best CRLs, ML Algorithms and ML models for each purpose of model building. Additionally, we utilised four SCA warning datasets to identify the performance patterns of ML models in relation to the characteristics of the datasets. However, existing studies discussed in Section 3 did not evaluate the model performance against the characteristics of the datasets. Furthermore, most of the existing studies discussed in Section 3, used one simple program slicing approach for every SCA warning type, such as using the issue line or using some lines above and below the issue line. In contrast, we employed multiple program slicing approaches, depending on the type of SCA warning. Consequently, this study makes a significant contribution to the body of scientific knowledge.

### 7.5 Implications for tool developers

SCA tools developers need to enhance SCA tools to reduce false-positive warnings. Therefore, they can employ ML models optimised with F1-Score performance metric to identify false-positive SCA warnings and detect common patterns of such warnings. Recognising these patterns will help tool developers decrease the occurrence of false-positives. Developers aim to identify all issues in source code, even if it results in false positive warnings. In contrast, users may prefer fewer false-positives and are less concerned about missing some real warnings. Consequently, they can implement a sensitivity parameter, allowing users to select the desired behaviour of the tools. This enables users to choose between reducing false-positives and maximising the detection of actual warnings.

### 7.6 Implications for the Practitioners community

ML models can be used to filter false positive SCA warnings. However, results generated from this study show that models do not perform perfectly. Therefore, classifying SCA warnings require ML models that are selected based on specific goals, as described in Section 2.2. For example, the practitioner community might need to maximise the overall model performance. Consequently, they can use ML models with the highest MCC values, which provide a model with the best overall performance. Additionally, MCC is a more robust performance metric when the dataset exhibits class-imbalance. On the other hand, they can use ML models with the best F1-Score values, if they need to correctly detect false-positive warnings as much as possible to minimise false warnings. Conversely, they can use ML models optimised with Balanced Accuracy metric if they need to correctly identify the maximum possible actual warnings with the minimum number of false-positive warnings. Finally, if they need to train and evaluate ML models quickly with limited computational resources, they can utilise models with the most efficient Computational Complexity. Therefore, they can fine-tune the models for better performance. Additionally, in this study, models are trained while optimising F1-Score. However, training can be done based on the purpose of model building.

Furthermore, results generated through this study show that the performance of the ML model depends on the CRL used, the ML algorithm employed, the characteristics of the dataset and the purpose of model building. Word2Vec and FastText can be used for SCA warnings, which involve a few nearby issue lines in the source code where there are no complex relationships between several source code regions. In contrast, when complex and semantic code representation is required to classify SCA warnings, Code2Vec, and CodeBERT can be utilised due to their ability to represent complex code effectively. The selection of the ML algorithm is based mostly on the complexity of the SCA warnings. If there are more false-positives in the source code, or the complexity of the SCA warning is higher, then deep learning ML algorithms can be used. On the other hand, if the warnings are simple, then using simpler models such as RF is sufficient. Therefore, the practitioner community can utilise CRLs and ML algorithms, based on the characteristics of the dataset, rather than relying on these technologies solely for novelty and advanced methodologies. Moreover, the use of performance metrics tailored to the task's purpose is also vital, rather than relying on common metrics such as Accuracy and F1-Score, which have been used in previous studies [6, 8, 13, 40]. Furthermore, employing hyperparameter optimisation methods such as Optuna, along with advanced program slicing approaches based on the SCA rule, will significantly enhance the model's performance.

### 7.7 Implications for the research community

Many existing studies primarily focus on experimenting with ML algorithms, overlooking the importance of evaluating CRLs for classifying false-positive SCA warnings [5, 6, 13]. However, we identified in this study that variants of both CRLs and ML Algorithms are essential. Therefore, it is vital to experiment with both CRLs and ML Algorithms to classify SCA warnings. Moreover, we identified that sometimes simple ML Algorithms, such as RF, perform well with the right CRL. It will be worth experimenting with multiple CRLs using simple ML Algorithms, such as RF and DTs, as these models have lower Computational Complexity. Therefore, it is faster to fine-tune these models.

Furthermore, we assessed our datasets, which were created by two SCA tools and exhibit different false-positive warning percentages. We found in our study that model performance depends on the traits of the datasets. Therefore, we urge academia to evaluate model performance using high-quality datasets and to make these datasets publicly accessible. Our previous research contributed to the academic community by making the datasets, which were used in this study, publicly available. Thus, the research community can assess and compare these results with other CRLs and ML Algorithms. Moreover, multiple performance metrics for ML model evaluation are emphasised. Additionally, it is vital to evaluate ML models based on the model-building goals as well as characteristics of the datasets and multiple performance metrics, as discussed in the Section 7, as finding the best ML model with the best Accuracy or F1-Score is not always enough for the practitioner community.

Moreover, it would be a good idea to train a small sample of the dataset to find the optimal combination of CRLs and ML algorithms and optimise hyperparameters, thereafter, training on the whole dataset to build the final model. This reduces the overhead of training multiple ML models with a large dataset. Ultimately, results show that a well-trained ML model can yield effective results even with a relatively small dataset. However, program slices of source code should be correctly extracted based on the warning type, and multiple technologies (CRL and ML model) need to be experimented with to find the best model for the dataset. Moreover, introducing more robust program slicing techniques based on the SCA rules/Warning types will contribute to academia.

## 8 THREATS TO VALIDITY

In this section, the potential threats to the validity of this study are discussed following previous guidelines [128].

**Internal Threats to Validity**

According to Croft et al. [129], the quality of the ML model depends on the quality of the dataset on which the ML model is built. In addition, the highest quality dataset can be created using manual annotation [130]. In our previous study, four manually annotated datasets were created and introduced in Section 4. The datasets were created by manual annotation according to defined standards. However, there may be incorrect annotations in those datasets due to errors during the annotation process (e.g., due to incomplete or misunderstood standards). To minimise errors, all authors discussed the standards, and several software tools were developed to assist in the annotation process. Additionally, all four datasets were rechecked at least once. Additionally, 10% of Dataset 1 and Dataset 3 were selected randomly, and two authors reviewed these samples together [131-133]. There were three data points out of 115 (51 in Dataset 1 and 64 in Dataset 3) that recorded disagreements at the beginning of the reliability assessment process. However, complete agreement on the 115 samples was reached after a round of discussions [134], attesting to the strong reliability of our datasets development [135].

The performance of ML models heavily relies on the hyperparameters used to train each ML model. Therefore, we used Optuna optimiser with TPE sampler to identify the best hyperparameters for each ML model. However, there is still a possibility that an alternative configuration could produce a better result. Moreover, there can be a bias in the results generated towards SCA tools. To mitigate that, we used two SCA tools. However, there can still be some bias, which we did not identify while analysing the results. This may affect the generalisability of the results for other SCA tools.

We used four CRLs: Word2Vec, Code2Vec, FastText, and CodeBERT. Code2Vec and CodeBERT are pre-trained with large sets of publicly available source code datasets. In contrast, Word2Vec and FastText are trained on non-source code text data. Therefore, following previous studies, we trained both Word2Vec and FastText with our two source code datasets (Stack Overflow dataset and Apache Tomcat dataset) [3, 5]. While training some CRLs with our source code datasets might bias the results in favour of those CRLs, we used the entire code dataset for training to minimise bias. Nonetheless, there remains a risk of producing biased results.

We used four datasets with sizes ranging from 117 to 298, which are relatively small compared to existing studies. Although we did not identify any overfitting issues in the results, there may still be some effect on the results.

While splitting the datasets into training and testing datasets, we did not consider the SCA rules of warnings. Consequently, for specific SCA rules, all warnings associated with the rule may be present solely in the training dataset, with none in the testing dataset, or vice versa. Therefore, the dataset will be significantly different based on the random seed used when splitting the dataset, which may cause different results to be generated by the ML model. However, we used the same random seed for each dataset to be consistent across all the ML models. Therefore, all the ML models that needed to be trained with the same dataset from our four datasets were trained and tested using the same portions of the dataset.

**External Threats to Validity**

There were warnings for a limited number of SCA rules in the datasets, which may cause generalisability issues. However, we did not filter any warnings based on the SCA rules, and we selected all the warnings which belonged to SCA rules that generated false-positive warnings. Therefore, the results should be relevant to any other dataset containing warnings generated for Java source code by PMD or SonarQube SCA tools.

We used CRLs and ML Algorithms that were the most effective in existing studies. However, there may be more advanced technologies that have not yet been evaluated and could be more effective than those we assessed in this study. To address this, we considered some recently introduced technologies that have not been extensively studied in existing studies, such as Code2Vec CRL and BiLSTM ML Algorithm.

**Conclusion Threats to Validity**

The conclusion made from this study is based on the defined methodology and the training datasets used. However, the conclusion might vary based on training datasets with different characteristics and alternative methods. To mitigate this, we used four datasets with various characteristics, and we used recent technologies/methodologies to evaluate ML models. These methodologies have been used in recent studies in a similar domain.

## 9 CONCLUSION AND FUTURE WORK

**Conclusion:** We conducted a comprehensive literature study to identify gaps in existing studies in identifying false-positives from SCA warnings. As a result, we identify several gaps in existing studies. One of the limitations is that existing studies do not comprehensively evaluate multiple CRLs and do not assess the impact of CRL performance on the overall ML model performance. Additionally, existing studies do not evaluate SCA warnings from multiple SCA tools to identify any possible performance patterns in the overall model performance changes based on the SCA tool used. Furthermore, they do not evaluate their ML models in multiple datasets with diverse characteristics. Moreover, existing studies do not consider model-building goals and multiple performance metrics to evaluate ML models under different model building goals, such as, maximising the overall model performance, correctly identifying false-positive warnings as much as possible to minimise false warnings, correctly detecting the maximum possible actual warnings with minimum false warnings, and faster training and evaluating ML models with limited computational resources so that models can be fine-tuned for better performance.

In our study, we fill these gaps through four research questions. RQ1 shows that the performance of the ML algorithm depends on the CRL used to represent the source code, and the best CRL for each ML algorithm is dependent on the performance metric used to evaluate the classification of SCA warnings. Moreover, RQ2 reveals that the performance of SCA warning classification depends on the ML algorithm used, and the best algorithm also varies based on the performance metric used to evaluate the warning classification. Furthermore, RQ3 shows that the performance of warning classification depends on the purpose of ML model building (related to performance metrics) and varies based on the utilised CRL, ML Algorithm, and the performance metric. Moreover, RQ4 revealed that the performance of warning classification is also dependent on the characteristics of the dataset used to train and test the model.

In this study, we evaluated multiple CRLs along with multiple ML algorithms and their individual impact on the overall performance of the model. In addition, we assessed these combinations of CRLs and ML Algorithms using numerous datasets with diverse characteristics in terms of various performance metrics. Additionally, these datasets were created by analysing two source code datasets using two SCA tools to identify any patterns in the used SCA tools and the overall performance of ML models. Overall, the best CRLs, ML Algorithms, and ML models (performance of false-positive warning classification) vary based on the characteristics of the training dataset and the goal of the model building. However, we did not identify any dependency on the SCA tools or the characteristics of the source code (code snippets or full-length source code) on the model performance. This is because of the use of advanced program slicing techniques, which adopt existing methodologies and customise them for each SCA warning rule.

The results generated in this study provide recommendations for SCA tool developers, the practitioner community, and the research community. Tool developers can train and use ML models to identify false positive warning patterns and reduce false-positive warnings. Additionally, introducing a sensitivity parameter can allow tool users to choose between having fewer false-positives or having actual warnings as much as possible. The practitioner community may select ML models based on their intention (refer to the purpose of building ML model in Section 2.2), as ML models do not produce perfect results. Additionally, training CRLs, such as Word2Vec and FastText, with source code that was analysed using the SCA tools, and training ML models multiple times, is vital to achieve the optimal ML model. Furthermore, the variant of CRL used to train the model affects the performance of the ML model. Therefore, it is emphasised that the research community should evaluate various CRLs as well as numerous ML algorithms. Additionally, the research community can consider factors such as Computational Complexity or interpretability in addition to other performance metrics such as MCC and F1-Score. Moreover, simple ML algorithms such as RF can at times also provide good performance compared to DL models. Therefore, it is necessary to evaluate simple models as well. Additionally, it is vital to report ML model performance in conjunction with the characteristics of the dataset, as the performance of the ML model is dependent on the characteristics of the dataset. Considering the ML model-building goals are also vital, as optimising ML models with commonly used performance metrics such as Accuracy, the F1-Score is not always applicable.

**Future Work:** In this study, we used several program slicing techniques. However, the quality of the program slice may affect the overall performance of the ML model. Therefore, we will conduct further research on the program slicing technique to enhance the overall performance of the model. Moreover, we employed machine learning models for this study to classify SCA warnings. However, we will research the possibility of using LLMs to enhance our outcomes, as utilising LLMs for a similar study may reduce the need for large training datasets.

## APPENDICES

### Appendix A Program Slicing Approaches for PMD Rules

| PMD Rules | Program_Slicing_Approach |
|---|---|
| AvoidArrayLoops | AST_FORWARD |
| AvoidDeeplyNestedIfStmts | AST_FORWARD_BACKWARD |
| AvoidThrowingNewInstanceOfSameException | DATA_DEPENDENCY |
| CloneMethodMustBePublic | AST_BACKWARD |
| CloneMethodMustImplementCloneable | AST_BACKWARD |
| CloseResource | DATA_DEPENDENCY |
| CommentRequired (Document comments) | ONE_LINE_UP |
| CommentRequired (Inline comments) | ISSUE_LINE_ONLY |
| DataClass | AST_FORWARD |
| DetachedTestCase | AST_FORWARD |
| DoNotUseThreads | ISSUE_LINE_ONLY |
| ForLoopCanBeForeach | DATA_DEPENDENCY |
| GuardLogStatement | ISSUE_LINE_ONLY |
| IdenticalCatchBranches | AST_FORWARD_BACKWARD |
| ImplicitSwitchFallThrough | AST_FORWARD |
| JUnit4TestShouldUseAfterAnnotation | AST_FORWARD_BACKWARD |
| JUnit4TestShouldUseTestAnnotation | AST_FORWARD_BACKWARD |
| JUnitTestsShouldIncludeAssert | AST_FORWARD |
| LawOfDemeter | DATA_DEPENDENCY |
| NonThreadSafeSingleton | DATA_DEPENDENCY |
| ProperLogger | ISSUE_LINE_ONLY |
| SingletonClassReturningNewInstance | AST_FORWARD |
| SingularField | AST_FORWARD_BACKWARD |
| TestClassWithoutTestCases | AST_FORWARD_BACKWARD |
| UnnecessaryLocalBeforeReturn | DATA_DEPENDENCY |
| UnusedAssignment | DATA_DEPENDENCY |
| UselessParentheses | ISSUE_LINE_ONLY |
| UseUtilityClass | AST_FORWARD |

# Appendix B Program Slicing Approaches for SonarQube Rules

| SonarQube Rule Code | Program_Slicing_Approach |
|---|---|
| java:S109 | ONE_LINE_UP |
| java:S135 | AST_FORWARD |
| java:S1067 | AST_BACKWARD |
| java:S1107 | ONE_LINE_DOWN |
| java:S1120 | AST_BACKWARD |
| java:S1182 | AST_FORWARD_BACKWARD |
| java:S1258 | AST_FORWARD |
| java:S1451 | TEN_LINE_DOWN |
| java:S1696 | ONE_LINE_DOWN |
| java:S1994 | ONE_LINE_DOWN |
| java:S2039 | ISSUE_LINE_ONLY |
| java:S2097 | AST_FORWARD |
| java:S2440 | AST_FORWARD |
| java:S2444 | AST_BACKWARD |
| java:S3012 | DATA_DEPENDENCY |
| java:S3047 | DATA_DEPENDENCY |
| java:S5738 | AST_BACKWARD |